\documentclass[a4paper,fleqn,usenatbib]{mnras}
\usepackage[T1]{fontenc}
\usepackage{ae,aecompl}
\usepackage{graphicx}
\usepackage{amsmath}
\usepackage{amssymb}

\title[LEAP: the large European array for pulsars]{LEAP: the large European array for pulsars}

\author[Bassa et al.]  {C.\,G.\,Bassa$^{1,2}$\thanks{email:
    bassa@astron.nl},
  G.\,H.\,Janssen$^{1,2}$,
  R.\,Karuppusamy$^{3,2}$,
  M.\,Kramer$^{3,2}$,
  \newauthor
  K.\,J.\,Lee$^{4,3,2}$,
  K.\,Liu$^{3,5,2}$,
  J.\,McKee$^{2}$,
  D.\,Perrodin$^{6,2}$,
  M.\,Purver$^2$,
  \newauthor
  S.\,Sanidas$^{7,2}$,
  R.\,Smits$^{1,2}$,
  B.\,W.\,Stappers$^2$\\
$^1$ASTRON, the Netherlands Institute for Radio Astronomy, Postbus 2, 7990 AA, Dwingeloo, The Netherlands \\
$^2$Jodrell Bank Centre for Astrophysics, The University of Manchester, Manchester, M13\,9PL, United Kingdom\\
$^3$Max Planck Institut f\"ur Radioastronomie, Auf dem H\"ugel 69, 53121 Bonn, Germany\\
$^4$Kavli institute for astronomy and astrophysics, Peking University, Beijing 100871, P.\ R.\ China\\
$^5$Station de Radioastronomie de Nan\c cay, Observatoire de Paris, 18330 Nan\c cay, France\\
$^6$INAF - Osservatorio Astronomico di Cagliari, via della Scienza 5, 09047 Selargius (CA), Italy\\
$^7$Anton Pannekoek Institute for Astronomy, University of Amsterdam, Science Park 904, 1098 XH Amsterdam, The Netherlands\\
}

\date{Accepted 2015 November 20.  Received 2015 November 20; in original form 2015 July 30.}
\pubyear{2015}

\begin{document}
\label{firstpage}

\pagerange{\pageref{firstpage}--\pageref{lastpage}} \pubyear{2013}
\maketitle

\begin{abstract}
The Large European Array for Pulsars (LEAP) is an experiment that
harvests the collective power of Europe's largest radio telescopes in
order to increase the sensitivity of high-precision pulsar timing. As
part of the ongoing effort of the European Pulsar Timing Array (EPTA),
LEAP aims to go beyond the sensitivity threshold needed to deliver the
first direct detection of gravitational waves. The five telescopes
presently included in LEAP are: the Effelsberg telescope, the Lovell
telescope at Jodrell Bank, the Nan\c cay radio telescope, the Sardinia
Radio Telescope and the Westerbork Synthesis Radio Telescope. Dual
polarization, Nyquist-sampled time-series of the incoming radio waves
are recorded and processed offline to form the coherent sum, resulting
in a tied-array telescope with an effective aperture equivalent to a
195\,-m diameter circular dish. All observations are performed using a
bandwidth of 128 MHz centered at a frequency of 1396\,MHz. In this
paper, we present the design of the LEAP experiment, the
instrumentation, the storage and transfer of data, and the processing
hardware and software. In particular, we present the software pipeline
that was designed to process the Nyquist-sampled time-series, measure
the phase and time delays between each individual telescope and a
reference telescope and apply these delays to form the tied-array
coherent addition. The pipeline includes polarization calibration and
interference mitigation. We also present the first results from LEAP
and demonstrate the resulting increase in sensitivity, which leads to
an improvement in the pulse arrival times.
\end{abstract}

\begin{keywords}
gravitational waves --- pulsars: general --- methods: data analysis --- techniques: interferometric
\end{keywords}

\section{Introduction}

Fundamental physics and our understanding of the Universe are at an
important crossroad. We can now compute the evolution of the Universe
back in time until a small fraction of a second after the Big Bang,
and the experimental evidence for our standard model of particle
physics has been exemplified by the detection of the Higgs boson
\citep{cks+12,aaa+12}.  At the centre of the theoretical understanding
of both of these branches of physics are Einstein's theory of general
relativity (GR) and the laws of quantum mechanics. Both theories are
extremely successful, having passed observational and experimental
tests with flying colours (e.g.\,\citealt{ksm+06}). Nevertheless, they
seem to be incompatible, and attempts to formulate a new theory of
quantum gravity, which would unite the classical world of gravitation
with the intricacies of quantum mechanics, remain an important
challenge.  In this quest it is therefore hugely important to know
whether GR is the right theory of gravity after all.

Because gravity is a rather weak force, it usually requires massive
astronomical bodies to test the predictions of Einstein's theory.  One
of these predictions involves the essential concept that space and
time are combined to form space-time that is curved in the presence of
mass.  As masses move and accelerate, ripples in space-time are
created that propagate through the Universe. These gravitational waves
(GWs) are known to exist from the observed decay of the orbital period
in compact systems of two orbiting stars as the GWs carry energy away
(e.g.\ \citealt{tw82,ksm+06}). After inferring their existence
indirectly in this way, the next great challenge is the {\em direct}
detection of GWs.

The frequency range for which we can expect GW emission from a variety
of sources covers more than 20 orders of magnitude.  Efforts to
measure the displacement of masses on Earth as GWs pass through
terrestrial laboratories are ongoing worldwide, with the operation and
upgrade of detectors such as (Advanced) LIGO \citep{aaa+09},
(Advanced) Virgo \citep{aaa+12b} or GEO600 \citep{glsc+10}. These
detectors probe GWs at kHz-frequencies and are therefore sensitive to
signals from merging binary neutron stars or black hole systems. At
slightly lower GW frequencies, a space-based interferometer like the
proposed eLISA observatory will be sensitive to Galactic binaries and
coalescing binary black holes with masses in the range of $10^4$ to
$10^6$\,M$_\odot$ \citep{aab+13}.

To reach a much lower GW frequency range (complementary to the
frequency range covered by ground-based detectors), we can use
observations of radio pulsars. Radio pulsars are spinning neutron
stars that emit beams of radio emission along their magnetic axes. The
pulses of radiation detected by radio telescopes correspond to the
passing of the narrow beam across the telescope with each rotation.
The fact that these pulses arrive with such regularity, from the best
pulsars, means that they act like cosmic clocks. In a Pulsar Timing
Array (PTA) experiment, we can use these most stable pulsars,
millisecond pulsars (MSPs), as the arms of a huge Galactic
gravitational wave detector, to enable a direct detection of GWs
\citep{det79,hd83}.

There are currently three major PTA experiments. In Australia, the
Parkes Pulsar Timing Array \citep{mhb+13} is utilising the 64-m Parkes
telescope. In North America, NANOGrav is making use of the 100-m Green
Bank Telescope (GBT) and the 305-m Arecibo telescope
\citep{dfg+13}. In Europe, the largest number of large radio
telescopes is available: the European Pulsar Timing Array (EPTA) has
access to the 100-m Effelsberg telescope in Germany, the 76-m Lovell
telescope at Jodrell Bank in the UK, the 94-m equivalent Westerbork
Synthesis Telescope (WSRT) in the Netherlands, the 94-m equivalent
Nan\c cay Radio Telescope (NRT) in France and, as the latest addition,
the 64-m Sardinia Radio Telescope (SRT) in Italy. For a recent summary
of the details of the mode of operation of the EPTA, its source list
and experimental achievements (e.g.\ the derived limits for the signal
strength of a stochastic gravitational wave background or the energy
scale of cosmic string networks) and major theoretical studies, we
refer to \citet{kc13,ltm+15}; Desvignes et al.\,(submitted). All three
experiments also work together within the International Pulsar Timing
Array (IPTA, \citealt{haa+10,man13}).

Despite the apparent simplicity of a PTA experiment, the timing
precision required for the detection of GWs is very much at the limit
of what is technically possible today. Indeed, all ongoing efforts
summarised above currently fail to achieve the needed sensitivity
\citep{dfg+13,ltm+15,src+13}. As timing precision increases
essentially with telescope sensitivity (up to a point where the
changing interstellar medium along the line-of-sight and the intrinsic
pulse jitter become dominant, e.g.\ \citealt{lvk+11,cs10}), an
increase in telescope sensitivity is needed. In the future, radio
astronomers expect to operate a new radio telescope known as the
Square-Kilometre-Array (SKA). The study of the low-frequency GW sky is
one of the major SKA Key Science Projects \citep{jhm+15}.  The SKA
sensitivity will be so large (ultimately up to two orders of magnitude
higher than that of the largest steerable dishes) that GW studies may
become routine and will open up an era of GW astronomy that will allow
us to study the universe in a completely different way.

In this paper, we present the first comprehensive introduction to the
Large European Array for Pulsars (LEAP), a new experiment that uses a
novel method and observing mode to harvest the collective power of
Europe's largest radio telescopes in order to obtain a ``leap'' in the
PTA sensitivity. The long-term aim for LEAP is to go beyond the
sensitivity threshold needed to obtain the first direct detection of
GWs. LEAP represents the next logical, intermediate step between the
current state-of-the-art of pulsar timing and the sensitivities
achievable with the SKA. The efforts and technical advances that LEAP
brings (as described below) are essential steps towards the
exploitation of the SKA and its study of the nHz-GW sky.

The LEAP experiment is introduced in \S\ref{sec:experiment}; in
\S\ref{sec:telescopes} we describe the participating telescopes and
the instruments; in \S\ref{sec:analysis} the pipelines involved in the
calibration and analysis of the data are explained. The observing
strategy is outlined in \S\ref{sec:observing_strategy} and initial
results are presented in \S\ref{sec:results}. We conclude in
\S\ref{sec:conclusions}.

\section{Experimental design}
\label{sec:experiment}


The goal of the LEAP project is to enhance the sensitivity of pulsar
timing observations by combining the signals of the five largest
European radio telescopes. The combination of individual telescope
signals can be done in two ways: coherently and incoherently.  In the
{\em incoherent} addition signals are added after detection (squaring
of the signal) hence removing the phase information of the
electromagnetic signal received by the individual telescopes, so that
the signal-to-noise ratio (S/N) increases with the square-root of the
number of added telescopes\footnote{In the case of telescopes with
  identical apertures and receivers, and uncorrelated noise.}. By
adapting proven techniques from existing Very Long Baseline
Interferometry (VLBI) experiments \citep[e.g.][]{tms91}, the phase
delays between the signals received at the individual telescopes can
be determined and corrected for, allowing for the {\em coherent}
addition of the signals (e.g.\ as described for LOFAR in
\citealt{sha+11}).  In this mode, the telescopes form a ``tied array''
beam that is pointed to a specific sky position (here that of a
millisecond pulsar). In the standard operation mode described below,
LEAP forms a single tied-array beam.  In this case, the S/N of the
LEAP observation is the (optimal) linear sum of the S/Ns of the
individual telescopes.

Forming the coherent LEAP tied-array beam shares many similarities
with a multi-element interferometer. In both cases, the individual
telescopes observe the same source over an identical range of
observing frequencies and correct the signals of the individual
telescopes for (differences in) the delays due to geometry,
atmosphere, instruments and clocks. In an interferometer, the
correlated signals are ultimately used to form images with high
spatial resolution, while for a tied-array, the signals from the
individual telescopes are added coherently in phase to form the
coherent sum. For short baselines of up to several kilometers, such as
for multi-element interferometers like the Australian Telescope
Compact Array (ATCA), the Jansky Very Large Array (JVLA), the Giant
Metre Radio Telescope (GMRT), the Low Frequency Array (LOFAR) and the
Westerbork Synthesis Radio Telescope (WSRT), these corrections can be
applied in analog or digital hardware, or software, producing the
tied-array signal in (or near) real-time
(e.g. \citealt{kss08,rbg12}). For longer baselines, it is usually
required to store the digitized Nyquist-sampled time-series and
process the data offline.  This approach is used in imaging
observations for long baseline interferometers such as global VLBI
observations or usually that of the European VLBI Network. Recent
progress with the new \textsc{SFXC} software correlator would allow
the formation of a tied-array out of the telescopes participating in
the European VLBI Network \citep{kk14,kkp+15}.

The LEAP project forms a tied-array telescope specifically designed to
provide high S/N observations of the MSPs that are in the EPTA (see
Table\,2 in \citealt{kc13}, and also Desvignes et al.~2015). Due to
the availability of sensitive L-band (1.4\,GHz) receivers at all EPTA
telescopes, LEAP observations are obtained at 1396\,MHz with an
overlapping bandwidth of 128\,MHz. During monthly observing sessions,
both pulsars and suitable phase calibrators are observed, and the data
are recorded to disk. These disks are then shipped to Jodrell Bank
Observatory, where the data are correlated (in order to determine the
relative phase delays) and coherently added using software running on
a high-performance computer cluster.

\section{Telescopes and instruments}
\label{sec:telescopes}

\subsection{Telescopes}
We describe here in more detail the telescopes presently involved in
LEAP:

The 100-m telescope located in Effelsberg, Germany, is a
fully-steerable parabolic dish with an altitude-azimuth mount, and is
operated by the Max-Planck Institut f\"ur Radioastronomie. For LEAP
observations, depending on scheduling constraints, one of the two
L-band (1.4\,GHz) receivers (multi-beam or single-pixel) is used. Both
receivers provide signals corresponding to the two hands of circular
polarization at their outputs. The receivers use cryogenically-cooled
low noise amplifiers (LNAs) based on high electron mobility
transistors (HEMT), resulting in a system temperature of 24\,K. At
L-band (1.4\,GHz), the telescope has a gain of 1.5\,K~Jy$^{-1}$.

The 250-foot (76.2-m) Lovell telescope at Jodrell Bank Observatory has
a parabolic surface with an altitude-azimuth mount. The telescope is
operated by the Jodrell Bank Centre for Astrophysics at the University
of Manchester.  A cryogenically-cooled receiver that is placed at the
primary focus and is capable of observing a 500 MHz wide band between
1.3 and 1.8\,GHz with a system temperature of 25\,K. This receiver has
linear feeds, but uses a quarter-wave plate to produce two hands of
circular polarization. The telescope gain for L-band (1.4\,GHz) observations is
1\,K~Jy$^{-1}$ at 45$^\circ$ of elevation.

The Nan\c cay radio telescope is a transit telescope of the Krauss
design, in which the radiation is reflected via a movable flat mirror
onto a spherical mirror, and then received at a movable focus cabin.
The telescope has an equivalent diameter of 94\,m. Depending on the
declination of the source, the telescope can track sources for
approximately 1\,hour. The L-band receiver covers the frequency range
from 1.1 to 1.9\,GHz with a system temperature of 35\,K, and has a
telescope gain of 1.4\,K~Jy$^{-1}$ at these frequencies.

The 64-m Sardinia Radio Telescope located in San Basilio, Sardinia, is
a fully-steerable parabolic dish with an altitude-azimuth mount and a
modern active surface that makes it one of the most
technologically-advanced telescopes in the world. It is the newest
addition to the LEAP project. The SRT joined LEAP in July 2013 during
its scientific validation phase. LEAP observations are done using a
cryogenically-cooled dual-band 1.4\,GHz and 350\,MHz confocal receiver
at the primary focus of the telescope. The L-band receiver has a
bandwidth of 500\,MHz (ranging from 1.3 to 1.8\,GHz), a system
temperature of 20\,K and has linear feeds. The corresponding telescope
gain is 0.63\,K~Jy$^{-1}$.

The Westerbork Synthesis Radio Telescope is an interferometer used as
a tied-array consisting of 14 equatorially-mounted, 25-m diameter,
fully-steerable parabolic dishes \citep{bh74}. The telescopes are
equipped with multi-frequency front-ends (MFFEs) that cover
frequencies from 110\,MHz to 9\,GHz in both polarizations almost
continuously. For LEAP observations, the MFFEs are tuned to receive
linearly polarized signals from eight overlapping 20\,MHz subbands
between 1.3 and 1.46\,GHz. The overlaps are necessary to match the
subbands generated by the other four LEAP telescopes. The subbands
from the 25-m telescopes are separately sampled at 2-bit resolution
and are then digitally combined in the tied-array adder module (TAAM),
after applying the appropriate geometric delay in each sampled subband
signal. This coherently-added signal is equivalent to the signal from
a 94-m diameter parabolic dish, and results in a system temperature of
27\,K and a telescope gain of 1.2\,K/Jy. Since the WSRT is currently
in the process of transitioning to the new APERTIF observing system
\citep{vov+08}, for LEAP observations we have used a varying number of
10 to 13 of the available 25-m dishes.

\subsection{Instruments}
To form the LEAP tied-array each observatory required an instrument
capable of recording Nyquist-sampled time-series over the LEAP
bandwidth. These time-series are typically referred to as baseband
data and represent the voltages measured at the telescope and sampled
at the Nyquist sampling rate. For the LEAP project, these baseband
recording instruments are required to sample the two polarizations of
the radio signal at 8-bit resolution over 128\,MHz of bandwidth, and
thus need to be capable of recording data at a rate of
4\,Gb\,s$^{-1}$. 

At the start of the project, VLBI baseband recording instruments were
available at Effelsberg, Jodrell Bank and WSRT. We decided not to use
those for LEAP as they use different signal chains compared to the
pulsar instruments in operation at those telescopes. Instead, we built
on our experience gained with the \textsc{PuMa\,II} instrument at WSRT
(see below), to design and build instruments for the other telescope
capable of recording baseband data. This approach allowed us to use
these instruments for regular/EPTA pulsar timing observations using
\textsc{DSPSR} \citep{sb10} to perform real-time coherent dedispersion
and folding. As such, instrumental time-offsets are minimized.

At WSRT, the TAAM generates Nyquist-sampled data of $8\times20$\,MHz
subbands at a resolution of 8\,bits. The \textsc{PuMa\,II} instrument
\citep{kss08} then records the baseband data onto disks attached to
separate storage nodes. At Nan\c cay, the \textsc{BON512} instrument
\citep{ctg+13} uses a ROACH FPGA board\footnote{Reconfigurable Open
  Architecture Computing Hardware (ROACH) FPGA board developed by the
  Collaboration for Astronomy Signal Processing and Electronics
  Research (CASPER) group; \texttt{http://casper.berkeley.edu/}} to
sample, digitise and polyphase-filter an input bandwidth of 512\,MHz
at 8\,bits into a flexible number of preset subbands. For standard
pulsar observations, the baseband data of each of these subbands are
sent over 10Gb Ethernet to processing nodes where GPUs perform
real-time coherent dedispersion and folding. For the LEAP project, the
disk space in one of the processing nodes was expanded to 55\,TB to
allow the baseband recording of $8\times16$\,MHz subbands.

At Effelsberg, Jodrell Bank and Sardinia, baseband recording
instruments were designed and built specifically for LEAP. These also
utilise a ROACH FPGA board where iADC analog-to-digital converters
perform the digitisation and Nyquist sampling of two polarizations at
8-bit resolution and for a bandwidth of up to 512\,MHz.  The ROACH
FPGA runs firmware based on the PASP\footnote{Packetised Astronomy
  Signal Processor library developed by the CASPER group.} library
blocks to perform a polyphase filterbank and generate subbands, which
are subsequently packetised as UDP packets and sent over the 10Gb
Ethernet network interfaces of the ROACH board. The UDP packets are
received by a cluster of computers where the baseband data are
recorded to disk using the \textsc{PSRDADA}
software\footnote{\texttt{http://psrdada.sourceforge.net/}}.  Absolute
timing is achieved by starting the streaming of data from the ROACH at
the rising edge of a one-pulse-per-second timing signal provided by
the observatory clocks. At the observatories, the ROACH iADC boards
are operated at clock speeds that fully sample the bandwidth provided
by the front-end, and produce at least 8 subbands with a bandwidth of
16\,MHz.  The analog signal chain at the observatories are set up so
that the center frequencies of these subbands are 1340, 1356, 1372,
1388, 1404, 1420, 1436 and 1452\,MHz, respectively.

The baseband data generated during LEAP observations from WSRT, Nan\c
cay, Effelsberg and Sardinia are sent to Jodrell Bank, where the
correlation and further processing is done on a dedicated computer
cluster, as described in Sects.\,\ref{sec:storage} and
\ref{sec:analysis}.

\subsection{Storage and processing hardware}
\label{sec:storage} To facilitate the storage and transfer of data
from the remote observatories to Jodrell Bank, storage computers with
removable disks were installed at Effelsberg, WSRT and
Sardinia. During LEAP observations, the raw baseband data of each
telescope are recorded onto the disks of the instrument. At the end of
the observing run, the data are transferred to the local storage
machine and the removable disks are then shipped to Jodrell Bank,
where they are placed into similar storage computers for offline
processing. After processing has finished, the removable disks are
shipped back to the remote observatories for re-use.

The baseband data obtained at Jodrell Bank are immediately transferred
over the internal network to one of the storage computers, while the
presence of a fast data-link between Nan\c cay and Jodrell Bank allows
the data obtained at Nan\c cay to be transferred directly over the
internet to one of the storage computers at Jodrell Bank.

At Jodrell Bank, a high performance computer cluster is used to
correlate and coherently add the baseband data from the individual
telescopes. The cluster consists of 40 nodes, each with two Quad core
Intel Xeon processors, 8\,GB of RAM and 2\,TB of storage.

\section{Data processing pipeline and calibration}
\label{sec:analysis}

A software correlator and beamformer were developed specifically for
the LEAP project to process the single-telescope baseband data and
form the coherent addition of these data. The correlator and
beamformer are part of a data processing pipeline that automates most
of the processing.

\begin{figure*}
  \includegraphics[width=\textwidth]{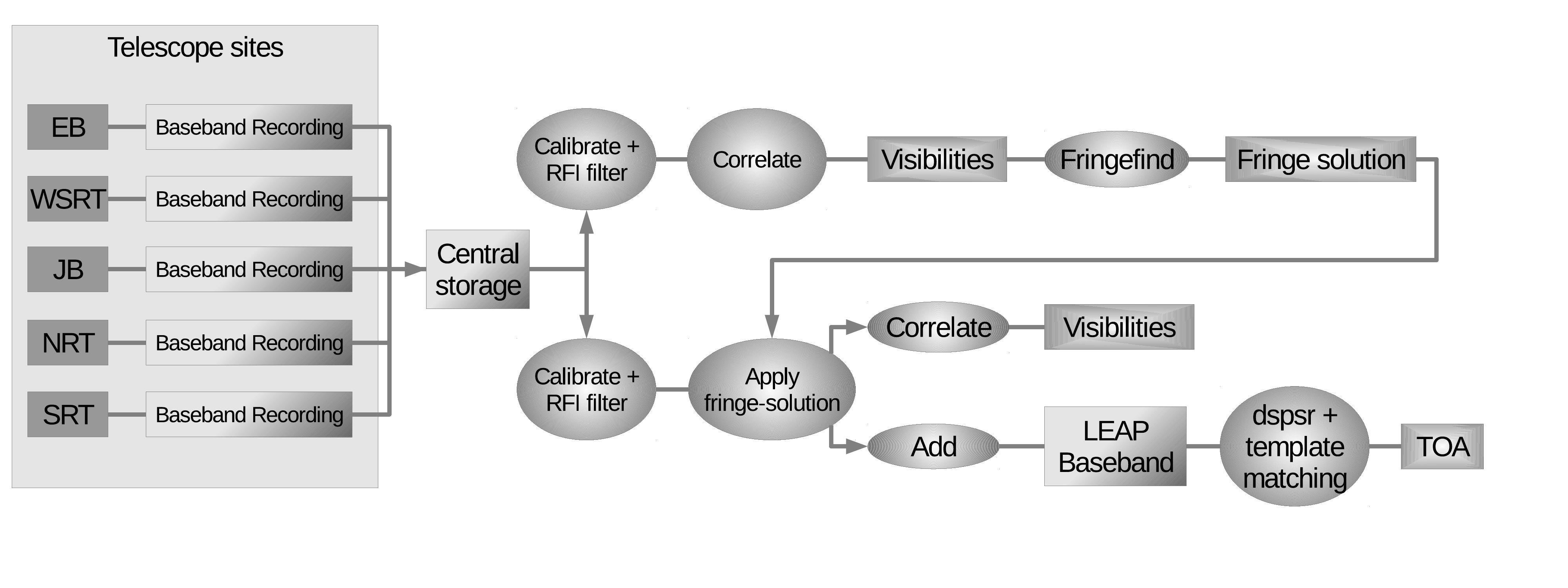}
  \caption{A flowchart of the LEAP data processing pipeline. Each
    observatory stores the baseband data from single-telescope LEAP
    observations on disk. The data are then transferred to the central
    storage machine at Jodrell Bank Observatory. There, polarization
    calibration and RFI mitigation filters are applied to the
    single-telescope data, which are then correlated, resulting into a
    fringe-solution for each of LEAP's baselines (ten telescope pairs
    in total). At this stage, we apply the fringe-solution to each
    telescope's baseband data (again after polarization calibration
    and RFI mitigation), correlate the time-series again, and check
    the resulting `visibilities' to verify that the fringe-solution is
    indeed correct.  The baseband data (to which the fringe-solution
    is applied) are then added together in phase, forming the LEAP
    tied-array. The added baseband data are processed as normal timing
    data. The data is then dedispersed and folded (using
    \textsc{DSPSR}) and template matching is performed to produce the
    final pulse times-of-arrival (TOAs).}
  \label{fig:pipeline}
\end{figure*}

\subsection{Data processing pipeline}
\label{sec:pipeline}

A flowchart of the LEAP processing pipeline is shown in
Fig.\,\ref{fig:pipeline}.  The processing starts once the baseband
data of each 16\,MHz subband from all LEAP telescopes from one of the
observing sessions are online at the central storage machine at
Jodrell Bank Observatory.

During the first processing stage, the data from each telescope are
correlated to find the exact time and phase offsets between the
telescopes.  This is achieved by first applying an initial time offset
corresponding to the geometric delay, the clock delay and the hardware
delay by simply shifting one of the time-series by an integer number
of samples with respect to the other. The remaining time delay is a
fraction of a time sample (see \S\,\ref{sec:phasecal}). The baseband
data are then Fourier-transformed (channelized) to the frequency
domain to form complex frequency channels. This is performed in
time segments of typically 100 samples, leading to 100 frequency
channels for each time-segment. The polyphase filters implemented in
the digital instruments at Effelsberg, Jodrell Bank, Nan\c cay and
Sardinia provide complex valued time-series, requiring the
complex-to-complex Fourier transform to channelize the data. In the
case of WSRT, real-valued time-series are created and the
real-to-complex Fourier transform is used to generate the channelized
complex time-series. When converted to the frequency domain, the
polarization is converted from linear to circular and the polarization
calibration is applied (\S\,\ref{sec:polcal}). At this stage the RFI
mitigation methods are also applied (\S\,\ref{sec:rfizap}). The
remaining fractional delay is corrected for by rotating the complex
values of each frequency channel in phase. The corresponding complex
time-series for each baseline pair and frequency channel are then
correlated to form 'visibilities'. As such, the correlator is of the
FX design, where the Fourier transform (F) is followed by the
correlation (X), similar to other software correlators like
\textsc{DiFX} \citep{dtbw07} and \textsc{SFXC} \citet{kkp+15}.

The visibilities are averaged in time, allowing the residual time and
phase offsets between each pair of telescopes to be extracted by
applying the global fringe fitting method from \citet{sc83}. An
initial Fourier transform method is used to find a fringe solution to
within one sample. This solution is then applied to a least-squares
algorithm that makes use of phase closure and involves minimizing the
difference between model phases and measured phases by solving for the
phase offset of each telescope (fringe phase), the time slope (fringe
delay) and the phase drift (fringe rate). The fits are performed
independently on both left-hand-circular and right-hand-circular
polarizations. The resulting fringe rates are averaged over both
polarizations.

During the second processing stage, the exact time and phase offsets
with respect to a reference telescope are applied to the baseband data
from each telescope. An amplitude scaling is also applied to these
data to ensure maximum sensitivity (see Sect.\,\ref{sec:ampcal}).

\subsection{Phase calibration and pulsar gating}
\label{sec:phasecal}
Creating the LEAP tied-array beam requires the baseband data from each
telescope to be corrected for an appropriate time delay and phase
shift before they can be added coherently. The time and phase delays
between the time-series from individual telescopes consists of four
components. First, the largest delays are due to differences in
geometry that result in different path lengths that the signal has to
travel. Second, there are differences between each observatory's local
clocks.  The third component consists of instrument-specific delays
due to cables and electronic components. Finally, the atmosphere (both
ionosphere and troposphere) introduces a delay as a time-varying
phase-shift of the radio-wavefront, which depends on the time-varying
conditions of the local atmosphere as well as the wavelengths of the
radio waves\footnote{The chosen observing frequency for LEAP of
  1.4\,GHz lies in a regime where both tropospheric and ionosperic
  effects are small.}.

The geometric delays can be largely corrected for by using the known
terrestrial positions of the telescopes, telescope pointing models and
celestial position of the source (calibrator or pulsar). The long
baselines in LEAP mean that our tied-array beam is very small and it
is therefore essential to have an accurate position for the right
epoch. It is therefore vital to include any known proper motion terms
when calculating the true position for the observing epoch. For LEAP,
these delays are calculated using the
\textsc{CALC}\footnote{\textsc{CALC} is part of the Mark-5 VLBI
  Analysis Software Calc/Solve} program \citep{rb80}. For our
pipeline, we make use of a C-based wrapper for \textsc{CALC}, which is
part of the \textsc{DiFX} software correlator \citep{dtbw07}. Applying
the geometric delays and clock delays requires a reference location
and a reference time standard. We have chosen to reference the time
series of the individual telescopes to the Effelsberg telescope. This
choice was made primarily because the Effelsberg telescope is the one
with the largest aperture. Because the time and phase delays are
determined on baselines that include Effelsberg, the corrections are
relative, not absolute. As a consequence, the corrected and
subsequently added baseband time-series can be treated for further
analysis as if they were observed by Effelsberg in terms of the
geometric delays and clock offsets normally used in pulsar timing.

\begin{figure}
  \includegraphics[angle=270,width=8cm]{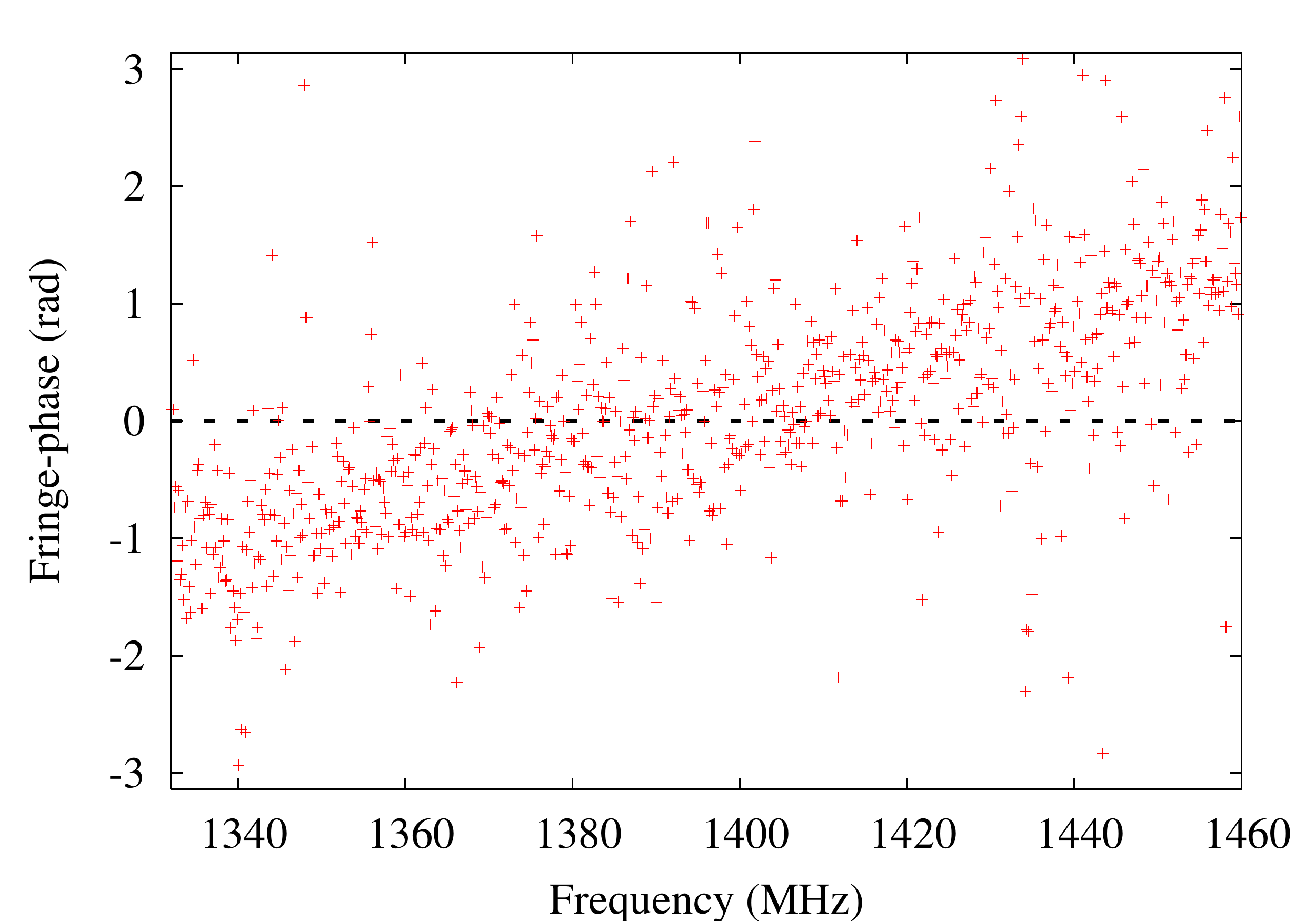}\\
  \includegraphics[angle=270,width=8cm]{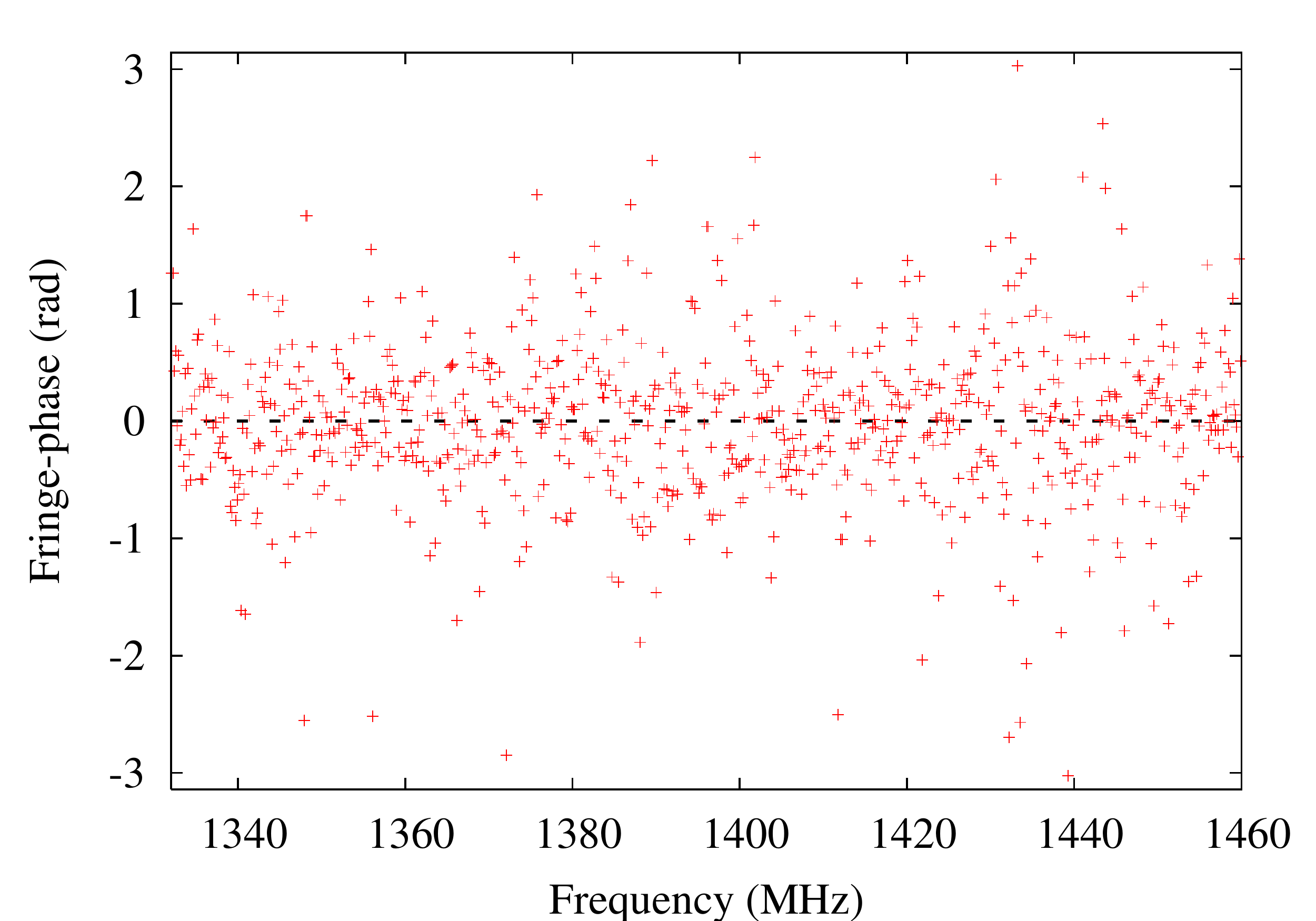}\\
  \caption{Fringe solution from a calibrator versus the fringe
    solution from the pulsar itself. These two panels show the
    visibility phase between the baseband time-series from Effelsberg
    and WSRT from the first 5\,minutes of an observation of
    PSR\,J1022+1001, taken on February\,24, 2015. The x-axis shows the
    observing frequency from 1332 to 1460\,MHz. The y-axis shows the
    visibility phase between the two time-series for each frequency
    channel (in units of radians). The top graph shows the visibility
    phase from the calibrator (taken 6\,min before the pulsar
    observation), applied to the pulsar observation. The bottom graph
    shows the fringe from the pulsar observation itself. A visibility
    phase of zero over the whole bandwidth means that the two signals
    are perfectly in phase and will thus add fully coherent. A
    residual time-offset between the two time-series will show up as a
    slope. The phase-calibrator is offset from the pulsar by $3^\circ$
    on the sky.}
  \label{fig:fringes}
\end{figure}

\begin{figure}
  \includegraphics[angle=270,width=8cm]{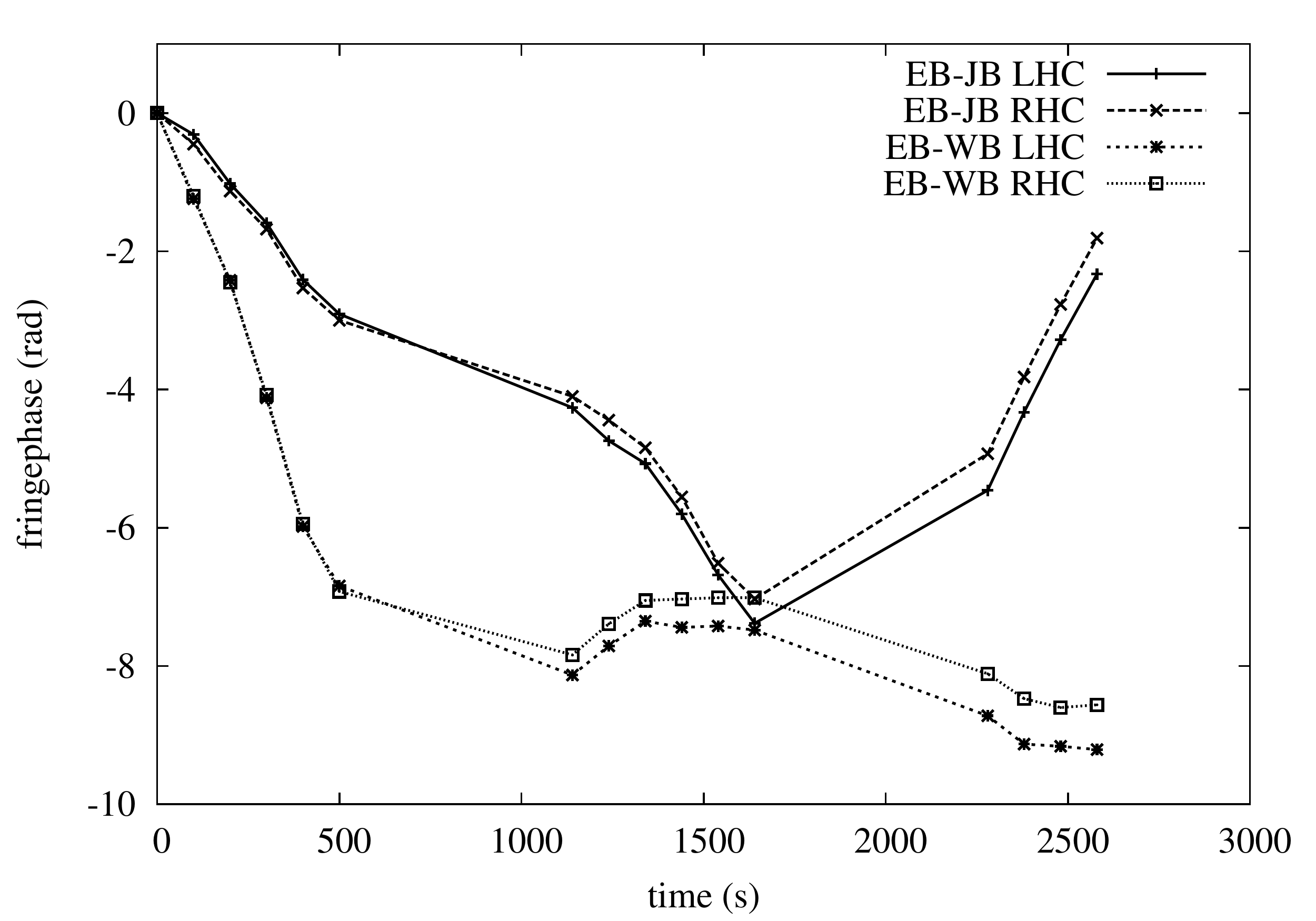}
  \caption{The evolution of the fringe-phase over time. The four lines show
    the drift in the fringe-phase in radians of a calibrator
    observation for the two baselines Effelsberg-Jodrell Bank and
    Effelsberg-WSRT for both polarizations: left-hand circular
    (LHC) and right-hand circular (RHC). It demonstrates that both the
    absolute value of the fringe-phase as well as the time-derivative
    of the fringe-phase (called fringe-drift) can change significantly
    on a timescale of minutes.}
  \label{fig:fringephase}
\end{figure}

The delays from the signal-path and the atmosphere are measured by
correlating the baseband data of the telescopes using the
purpose-built LEAP software. An initial fringe-solution of the
residual time and phase differences between each pair of telescopes is
found by correlating a calibrator source. However, the calibrator
source is typically offset by about 5$^\circ$ from the pulsar and
separated in time by several minutes. Because of this, the conditions
of the ionosphere/troposphere for the calibrator observation will be
different than for the pulsar observation, leading to a different
fringe-solution. Thus, when the fringe-solution from the calibrator is
applied to the pulsar data, it does not yield perfect coherence (see
Fig.\,\ref{fig:fringes}). In addition, the conditions of the
ionosphere/troposphere can change unpredictably on a timescale of
minutes, as shown in Fig.~\ref{fig:fringephase}. This means that the
observation would need to be interrupted to observe the calibrator at
least once every 15 minutes (or even every 5 minutes in case the
ionospheric conditions are very poor). As part of the processing
pipeline, we therefore developed a procedure to allow the
phase-calibration to be performed on the pulsar signal itself. This
method of calibrating on the target is called self-calibration and
widely used in interferometry.

To do this, we implemented a pulse binning technique to optimize the
sensitivity.  The visibilities within each individual pulse are
integrated into bins with a size equal to a fraction of the pulse
period.  This is done for each frequency channel. The bins from each
individual pulse are then added (folded) to the corresponding bins
from all previous pulses, using TEMPO to predict the exact pulse
period. A time-shift is applied to each individual channel to correct
for the dispersion delay. This results in average visibilities for
each pulsar phase bin, for each frequency channel and for each
baseline. Finally, the bins containing the on-pulse signal are
selected (this is the process of gating) and averaged together. This
yields visibilities for each baseline where only the on-pulse signal
of the pulsar contributes, and increases the signal-to-noise ratio
roughly by a factor equal to the reciprocal of the square-root of the
duty cycle. This procedure allows the fringes to be tracked over time
on the pulsar signal itself as the conditions of the
ionosphere/troposphere change, removing the need to switch between
pulsar and calibrator during the observation. Phase calibrating on the
target source uses the positional information of the pulsar, and hence
this approach can not be used for astrometry.

Once the total time and phase delays for each telescope with respect
to the reference telescope have been determined, they are applied to
the raw data in two stages. First, the baseband data from each
telescope are aligned to the nearest integer sample (62.5\,ns for a
complex sampled subband of 16\,MHz). The remaining fractional time
delay (a fraction of a sample) plus the measured delay in phase, is
corrected for by phase rotating the complex values of the channelized
time-series. After these corrections, the channelized time-series from
each telescope correspond in both time and phase with the time-series
from the reference telescope. These channelized time-series can thus
be added together coherently.

Finding a fringe-solution after correlating the time-series from the
telescopes can be impeded by a lack of pulsar signal, rapidly changing
conditions of the ionosphere/troposhere, extreme cases of RFI, or --
in the case of Nan\c cay -- by an irregular clock-drift\footnote{The
  rubidium clock that is providing the timing signals for the LEAP
  pulsar backend has typically two correction values per day with
  respect to the time standard at Paris-Meudon Observatory. The clock
  drift can be as large as 10\,ns within one hour, and can sometimes
  deviate from linear drift.}. In those instances where no
fringe-solution can be obtained, the time-series are added
\textit{incoherently}. The time series are then corrected for the
known time-delays by applying the geometric delay correction, the
clock correction, the instrumental delays and the fringe-solution from
the calibrator, which aligns the signals to within a few tens of
ns. Once the signals are time-aligned, they are added without
consideration of the relative phase of the electromagnetic signal
received by the individual telescopes. This is achieved by simply
adding the power of the baseband data. For incoherent addition, the
signal-to-noise ratio increases with the square-root of the number of
added telescopes\footnote{In the case of telescopes with identical
  apertures and receivers, and uncorrelated noise.}.

\subsection{Polarization calibration}
\label{sec:polcal}
\begin{figure*}
  \includegraphics[width=\textwidth]{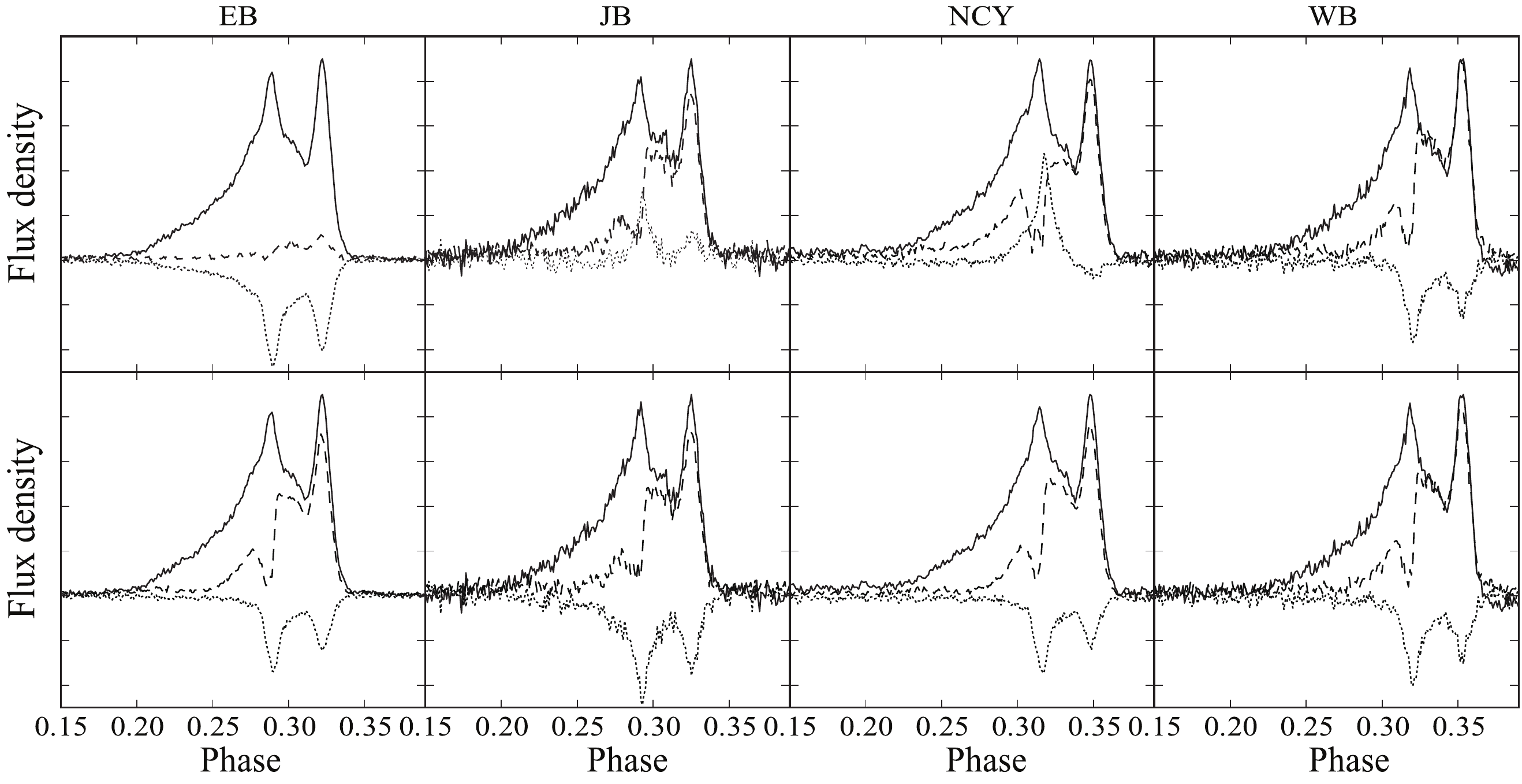}
  \caption{Pulsar profiles of PSR\,J1022+1001 as observed with the
    individual telescopes before and after polarization
    calibration. The solid, dashed, and dotted curves, are for total
    intensity, linear polarization, and circular polarization
    respectively. The top row are the profile without calibration, and
    the bottom row are the calibrated ones. The EB, JB, NCY, and WB
    abbreviations indicate the Effelsberg, Jodrell, Nan\c cay, and
    Westerbork telescopes. Here the y-axis, flux, takes an arbitrary
    unit, and x-axis is pulse phase. The calibrated profiles clearly
    show much better consistency.}
  \label{fig:pcal_comparison}
\end{figure*}

To maximize the coherency of the tied-array beam, it is crucial to
perform accurate polarization calibration that removes the effects
introduced by the telescope, receiver and instrument. This is
particularly important for LEAP, as each of the individual telescopes
is of a different design, uses different receivers and feeds, and we
are observing pulsars for which parts of the average pulse profiles
are up to 100\%-polarized. In Fig.\,\ref{fig:pcal_comparison} we
compare uncalibrated pulse profiles with profiles after calibration
using the method described below.

Here we briefly describe the LEAP polarization calibration scheme,
the details of which will be presented in a forthcoming paper. In
LEAP, polarization calibration is performed for each telescope
independently, before correlating and finding the fringes.  Performing
polarization calibration has two major benefits. First, it helps to
improve the S/N of fringe solutions, i.e.\ to determine accurate phase
offsets between telescopes. Second, performing polarization
calibration after coherent addition is complicated, since extra phases
have been introduced in the addition process. In fact, the expected
S/N of an un-calibrated fringe will be 22\% lower than the calibrated
one, assuming random differential phase between the two hands of
polarization and a 100\% polarized signal. It is thus hard to evaluate
the polarization performance of each telescope, and check the data
integrity individually.

For single telescope systems, the distortion of polarization can be
described by seven \emph{system parameters}\footnote{The $2\times2$
  complex Jones matrix has 8 real parameters that are required to
  specify it. However, the total phase shift is determined by fringe
  fitting, so only 7 parameters are required. The number of parameters
  can be reduced to 6 if one is not interested in the gain
  calibration.}. For a quasi-monochromatic wave, there are two major
parametrization schemes. In Britton's scheme \citep{Bri00}, there are:
the total gain, spinor transformation axes (four parameters) and the
transformation rotation angles (two parameters). In Hamaker's scheme
\citep{hbs96}, there are: the total gain, the gain-phase imbalance
(two parameters), leakage amplitude and phase (four parameters). The
two descriptions are equivalent. We adopt the Hamaker scheme in the
LEAP pipeline, however we do not assume that the polarization
distortions are small, since we are working with an inhomogeneous
array.

The aforementioned system parameters can be measured by comparing the
observed full-Stokes pulsar pulse profile to the standard profile
templates. The standard $\chi^2$ fitting minimizing the differences
between the template and the modeled profile is used to fit for the
system parameters of each frequency channel. In this way, the pulsar
itself is also used as the polarization calibrator in our
observations. PSR\,J1022+1001 and/or PSR\,B1933+16, for which the
pulse profiles show significant amounts of both linear and circular
polarization components, are normally used for polarization
calibration. Our approach is similar to the matrix template matching
method by \citet{str06}, except that we calibrate baseband data
directly.

\begin{figure}
  \includegraphics[width=8cm]{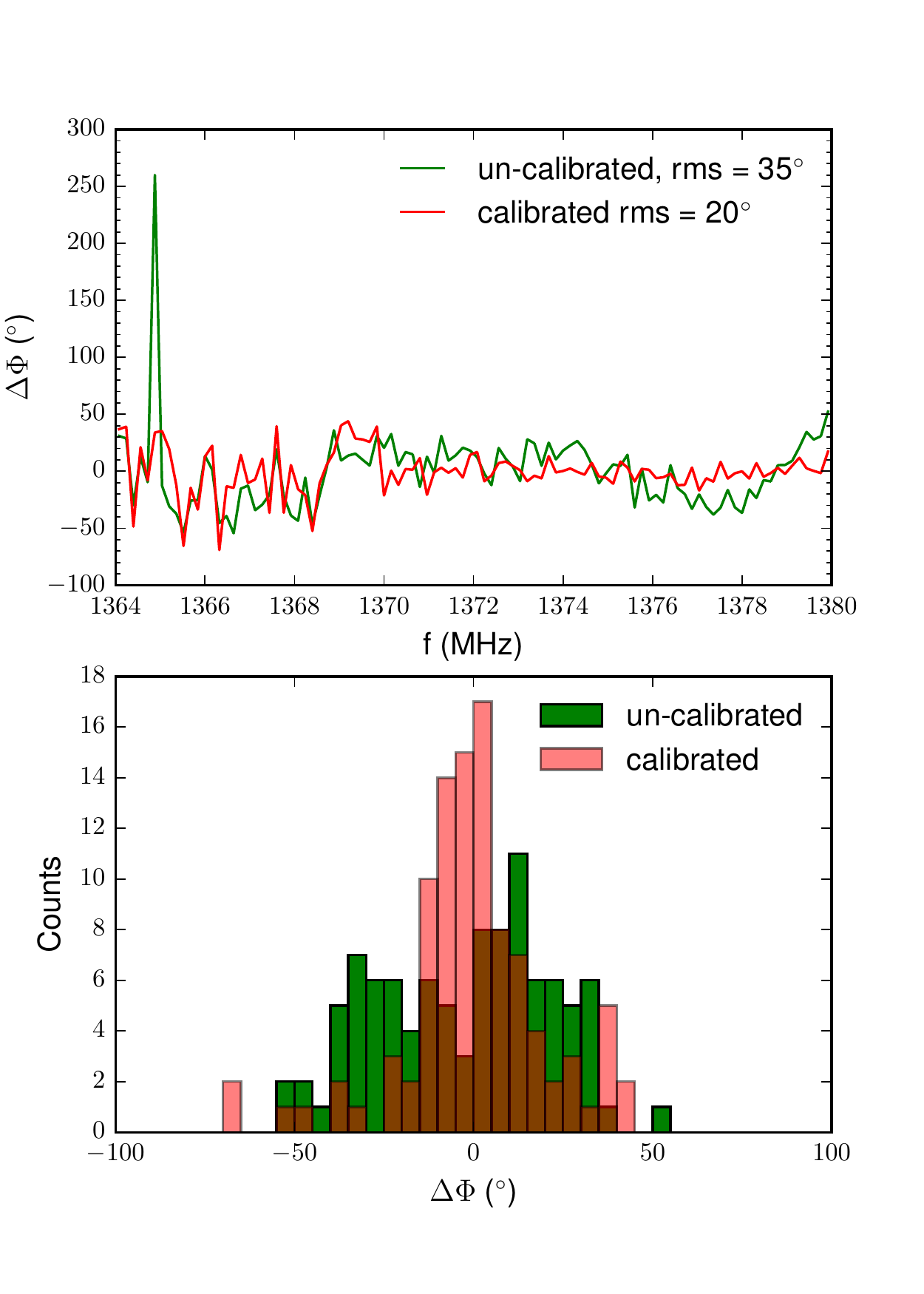}
  \caption{This figure illustrates the effects of the polarization
    calibration for a 10-second integration of a 16\,MHz subband of
    the Effelsberg-Nan\c cay baseline. The top panel shows the
    visibility phase $\Delta \Phi$ as a function of frequency with and
    without applying of the polarization calibration. A histogram of
    these phase delays with and without applying the polarization calibration
    is shown in the bottom panel. For this example, the average S/N
    of the visibilities shows an 18\% increase after polarization
    calibration, and the corresponding phase error is reduced by 40\%,
    i.e.\ the rms level of the visibility phase reduced is from
    $35^\circ$ to $20^\circ$. }
  \label{fig:pcal}
\end{figure}

There are three major steps in our algorithm. First, the observed
pulse profile is aligned with a template profile (using the algorithm
of \citealt{tay92}). Next, non-linear $\chi^2$-fitting is used to
derive the system parameters. These system parameters are then applied
to the observed profile in order to estimate the post-calibrated
profile. These steps are repeated until the solution converges, that
is when the fractional changes of the system parameters are smaller
than $10^{-7}$.  Our results show that the above iteration converges
most of the time, and that we can measure both the system parameters
and the phase offsets between the template and measured pulse profile
at the same time. This procedure is similar to using a noise diode as
a calibrator. However, because of the change of polarization angle
across the pulse profile, we are no longer limited to the case of
single-axial calibration, and are able to fix the whole set of system
parameters, including leakage terms. Indeed, we need to include such
terms to fully calibrate the Nan\c cay data. Figure\,\ref{fig:pcal}
shows the improvement in visibility phases after calibrating the
polarization.

\subsection{Amplitude calibration}\label{sec:ampcal}
To ensure maximum S/N of the added data, we have to apply an
appropriate weight to the baseband data from each of the telescopes,
where we have to consider that the final added data are written as
8-bit samples. To achieve this, we select a reference telescope and
measure the noise-levels from the baseband data from each telescope
and set the weights such that all samples are scaled to the
noise-levels of the reference telescope. We then take the S/N from the
average intensity profiles from the individual telescopes and scale the weights
with an additional factor given by:
\[
W_\mathrm{tel}=\sqrt{\frac{\mathrm{S/N}_\mathrm{tel}}{\mathrm{S/N}_\mathrm{ref}}},
\]
where $\mathrm{S/N}_\mathrm{tel}$ is the S/N of the telescope and
$\mathrm{S/N}_\mathrm{ref}$ is the S/N of the reference
telescope. This ratio of the S/N includes the telescopes' system
temperature relative to that of the reference telescope. The voltage
samples from each of the telescopes are then multiplied by the
corresponding weight before the addition, which maximizes the S/N of
the added data. At this stage, the samples are floating point
numbers. After the addition, a final scaling is applied such that the
standard deviation of the samples becomes one third of the dynamic
range of 8-bit data. This ensures minimal clipping and optimal use of
the dynamic range when the data is converted to 8-bit and written to
disk.

\subsection{Interference mitigation}\label{sec:rfizap}
In case of significant radio frequency interference (RFI), we have
implemented two methods to clean the data. The RFI mitigation step is
optional and performed right after the calibration. These RFI
mitigation methods are applied to the channelized data from the
individual telescopes before coherent addition.

The first form of RFI-mitigation consists of selecting and masking
frequency channels that contain narrow-band RFI. These channels are
selected via a simple algorithm that looks for channels with an
integrated power exceeding either a given threshold or deviating
significantly from its neighbors. These channels are then masked by
replacing the content with Gaussian noise with mean and rms determined
from neighboring time samples.

A second technique can be applied to data containing time-varying RFI,
or broadband RFI. This technique implements the method of spectral
kurtosis \citep{ng10a,ng10b} to remove RFI from some observations. It
provides unbiased RFI removal with a resolution of 6.25\,ms in time
and 0.16\,MHz in frequency. In each frequency channel and at each
telescope, the distribution of a time-series of 1000 samples of total
power is assessed for similarity to that expected from
Gaussian-distributed amplitudes. This is done using an estimator that
measures the variance divided by the square of the mean for these
power samples. When the power is derived from Gaussian amplitudes of
zero mean, the estimator has a probability density function (PDF) that
is independent of the variance of those amplitudes. It can therefore
be used to distinguish RFI on the premise that non-Gaussian amplitudes
are caused by RFI. The PDF is used to determine three-sigma limits for
the estimator, and a block of 2000 amplitude samples (1000 in each
polarization channel) is masked if it gives an estimator value outside
these limits. This excludes $0.27\%$ of RFI-free data, while excluding
most RFI-contaminated data.  The amplitudes of RFI-contaminated
samples are replaced by artificial Gaussian noise with the same
variance as nearby samples, in order to maintain a constant noise
level in the correlated amplitudes regardless of the number of
telescopes contributing to each sample. As before, the masked data is
replaced by Gaussian noise.

\begin{figure}
  \includegraphics[angle=270,width=8cm]{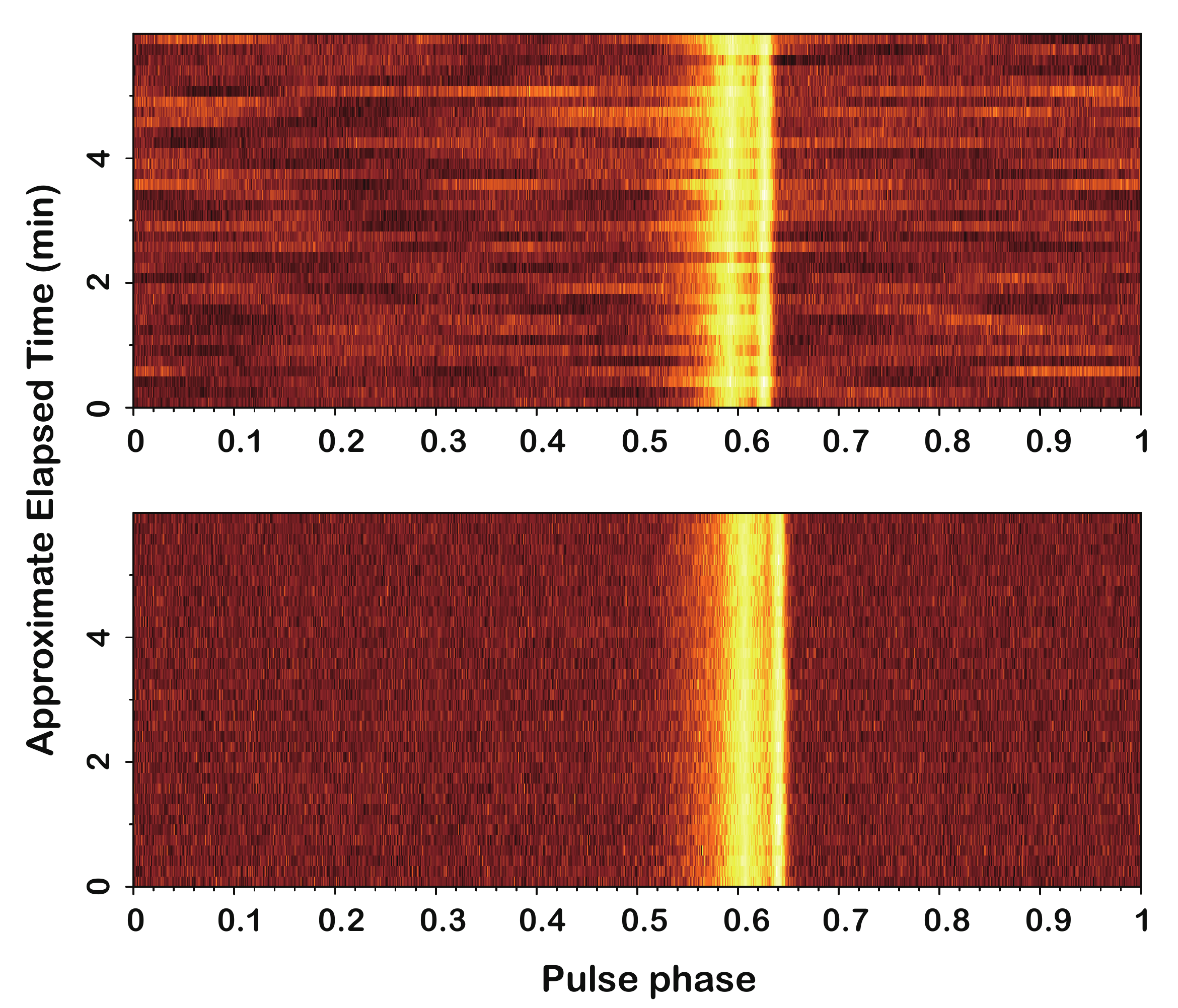}
  \caption{Pulsar phase-vs-time plot of coherently-added LEAP
    data of PSR\,J1022+1001, without (top) and with the spectral
    kurtosis RFI mitigation method (bottom). The observation was taken
    on July 27, 2013 with Effelsberg, Jodrell Bank, Nan\c cay, and
    WSRT. There was significant broadband RFI from the Nan\c cay
    observation, which dramatically changed the baseline of the
    coherently-added integration profile, as shown in the top
    panel. After applying the filter to Nan\c cay data only, the
    resulting LEAP data are significantly improved. }
  \label{fig:rfimitigation}
\end{figure}

We cannot generally define the percentage of RFI-contaminated data
that is excluded, because we do not know, a priori, the PDF of the
estimator derived from these data. Some RFI-contaminated data may not
be excluded if their PDF closely mimics that of Gaussian
amplitudes. However, our practical application has shown it to be
effective in automatically removing the vast majority of the dominant
RFI that would otherwise spoil our correlations (see
Fig.~\ref{fig:rfimitigation}). It is also possible that a very strong
pulsar signal could be misinterpreted as RFI by the spectral kurtosis
method but that does not happen when using the time and frequency
resolutions employed by LEAP.

\begin{table*}
  \caption{Pulsars and calibrators observed for the LEAP
    project. Notes: {\it a}) PSR\,J1518$+$4904 cannot be observed
    simultaneously with all five telescopes, therefore the Jodrell Bank,
    Effelsberg, and Sardinia telescopes observe PSR\,J1738$+$0333
    instead. {\it b}) PSR\,B1933$+$16 is used for polarization
    calibration as explained in Sect.\,\ref{sec:polcal} and is not
    included in the PTA list. Telescope codes: {\it E}: Effelsberg;
    {\it J}: Jodrell Bank; {\it N}: Nan\c cay; {\it S}: Sardinia; {\it
      W}: WSRT.}
  \begin{tabular}{llrl|llrl}
    \hline\hline
    Pulsar & Calibrator & length (min) & telescopes & Pulsar & Calibrator & length (min) & telescopes \\
    \hline
    & J0029+0554 & 3 & EJNSW &    & J1719+0817 & 3 & EJNSW \\
    J0030+0451 & & 40 & EJNSW &   J1713+0747 & & 50 & EJNSW \\
    & J0037+0808 & 3 & EJNSW &    & J1719+0817 & 3 & EJNSW \\[0.25em]
    & J0606$-$0024 & 3 & EJNSW &    & J1740+0311 & 3 & EJS \\
    J0613$-$0200 & & 60 & EJNSW &   J1738+0333$^a$ & & 60 & EJS \\
    & J0616$-$0306 & 3 & EJNSW &    & J1740+0311 & 3 & EJS \\[0.25em]
    & J0619+0736 & 3 & EJSW &     & J1740$-$0811 & 3 & EJNSW \\
    J0621+1002 & & 45 & EJSW &    J1744$-$1134 & & 45 & EJNSW \\
    & J0619+0736 & 3 & EJSW &     & J1752$-$1011 & 3 & EJNSW \\[0.25em]
    & J0743+1714 & 3 & EJSW &     & J1821$-$0502 & 3 & EJNSW \\
    J0751+1807 & & 40 & EJSW &    J1832$-$0836 & & 35 & EJNSW \\
    & J0743+1714 & 3 & EJSW &     & J1832$-$1035 & 3 & EJNSW \\[0.25em]
    & J0927$-$2034 & 3 & EJNSW &    & J1847+0810 & 3 & EJNSW \\
    J0931$-$1902 & & 40 & EJNSW &   B1855+09 & & 50 & EJNSW \\
    & J0932$-$2016 & 3 & EJNSW &    & J1847+0810 & 3 & EJNSW \\[0.25em]
    & J0957+5522 & 3 & EJSW &    & J1926$-$1005 & 3 & EJSW \\
    J1012+5307 & & 45 & EJSW &    J1918$-$0642 & & 20 & EJSW \\
    & J0957+5522 & 3 & EJSW &     & J1926$-$1005 & 3 & EJSW \\[0.25em]
    & J1015+1227 & 3 & EJNSW &   & B1933+16$^b$ & 5 & EJNSW \\
    J1022+1001 & & 45 & EJNSW &   B1937+21 & & 45 & EJNSW \\
    & J1025+1253 & 3 & EJNSW &    & J1946+2300 & 3 & EJNSW \\[0.25em]
    & J1028$-$0844 & 3 & EJSW &     & J2006$-$1222 & 3 & EJNSW \\
    J1024$-$0719 & & 45 & EJSW &    J2010$-$1323 & & 55 & EJNSW \\
    & J1028$-$0844 & 3 & EJSW &     & J2011$-$1546 & 3 & EJNSW \\[0.25em]
    & J1506+4933 & 3 & NW &       & J2130$-$0927 & 3 & EJNSW \\
    J1518+4904$^a$ & & 60 & NW &  J2145$-$0750 & & 45 & EJNSW \\
    & J1535+4957 & 3 & NW &       & J2155$-$1139 & 3 & EJNSW \\[0.25em]
    & J1554$-$2704 & 3 & EJNSW &    & J2232+1143 & 3 & EJNSW \\
    J1600$-$3053 & & 60 & EJNSW &   J2234+0944 & & 35 & EJNSW \\
    & J1607$-$3331 & 3 & JNSW &     & J2241+0953 & 3 & EJNSW \\[0.25em]
    & J1641+2257 & 3 & EJSW &     & J2303+1431 & 3 & EJNSW \\
    J1640+2224 & & 50 & EJSW &    J2317+1439 & & 40 & EJNSW \\
    & J1641+2257 & 3 & EJSW &     & J2327+1524 & 3 & EJNSW \\[0.25em]
    & J1638$-$1415 & 3 & EJNSW &   & & & \\
    J1643$-$1224 & & 35 & EJNSW &   & & & \\
    & J1638$-$1415 & 3 & EJNSW &    & & & \\
    \hline
  \end{tabular}
  \label{tab:srcs}
\end{table*}

\section{Observing strategy}
\label{sec:observing_strategy}

LEAP observations are crucial in that they complement the regular,
more frequent multi-frequency observations of the EPTA by adding
time-of-arrival measurements with the highest possible
precision. Observing sessions for LEAP are scheduled with an
approximately monthly cadence, each session lasting a minimum of
24\,hours. During each observing session, a set of millisecond pulsars
and phase calibrators are observed simultaneously with each of the
five radio telescopes. Since the first observations of June 2010, the
observing time per session, number of pulsars per session, and number
of participating telescopes per session have steadily increased.

Initial testing to aid in software development used eight of the
single 25-m WSRT dishes, obtaining 20\,MHz of bandwidth for a set of
6 millisecond pulsars. These data were used to test software
beamforming and allow a comparison with the output of the WSRT
hardware beamformer. The first long-baseline observations were
obtained in June 2011 using WSRT and Effelsberg. These observations
initially used five subbands of 20\,MHz, but switched to the
$8\times16$\,MHz setup starting in February 2012, when the Lovell
telescope at Jodrell Bank was included in the LEAP array. The Nan\c
cay telescope first joined in May 2012, initially with
$4\times16$\,MHz subbands, and since December 2012 with the full
128\,MHz bandwidth. Test observations with the SRT were obtained in
July 2013 for one 16\,MHz subband. Tests with one subband were then
performed monthly until January 2014. Finally, thanks to the
successful installation of an 8-node computer cluster, the telescope
joined full-length and full-bandwidth LEAP sessions in March 2014.

Through a memorandum of understanding between the participating
telescopes and institutes, observing time at Jodrell Bank, Nan\c cay
and SRT is guaranteed, while for Effelsberg and WSRT, the observing
runs are proposed through the peer-review process at these telescopes.
The long-term scheduling at Effelsberg and WSRT thus guides the
scheduling of the LEAP observing sessions, which are matched by the
Lovell, Nan\c cay and Sardinia telescopes.

Besides the principal requirement that the observed sources be
simultaneously visible from all sites, the observing schedule takes
the individual telescope constraints into account for each LEAP
session.  The primary observing constraint is set by the transit
design of Nan\c cay, where sources are visible for 60 to 90\,minutes
around culmination, depending on the declination of the source.  The
altitude-azimuth mounts of the Effelsberg, Lovell and Sardinia
telescopes usually do not allow observations at very small local
zenith-angles (i.e.\ LEAP observations avoid zenith angles of less
than 10$^\circ$), and have slew limits at certain azimuths related to
cable wrapping.  The equatorial design of WSRT limits observations to
hour angles from $-6$ to $+6$\,h around transit for each source.
Furthermore, WSRT requires a 3-min initialization time between
observations to configure the tied-array. This initialization time
overlaps with the slewing time for all telescopes, as well as with a
minimum observing length requirement of 6\,min for all observations
done with the Lovell Telescope.  The slewing rates, minimum observing
time and initialization time mostly impact the calibrator observations
before and after each pulsar observation, which are generally only
three minutes long. To obtain the most efficient overall observing
schedule and a maximum overlap between all telescopes for each
observation, LEAP requires all observations to end at the same time.

Besides the telescope constraints, the visibility of MSPs suitable for
pulsar timing array experiments also provides a stringent constraint
on the schedule. To first order, the most suitable pulsars are
clustered towards the inner Galactic plane, with very few pulsars at
right ascensions between $01^\mathrm{h}$ and
$05^\mathrm{h}$. Furthermore, to maximize the number of sources that
are visible at Nan\c cay, it is beneficial to include sources
separated equally in right ascension. To maximize the number of
suitable MSPs observable by LEAP, we moved away from continuous
24-hour observing sessions. Since the spring of 2013, we observe in
two sessions, spanning right ascension ranges from
$06^\mathrm{h}00^\mathrm{m}$ to $01^\mathrm{h}30^\mathrm{m}$ and
$15^\mathrm{h}30^\mathrm{m}$ to $21^\mathrm{h}00^\mathrm{m}$. The two
parts of a full LEAP run are usually separated by only a
day. Table\,\ref{tab:srcs} lists the pulsars and phase calibrators
observed by LEAP. The current selection of pulsars is based on an
optimization of using the best pulsars observed by the EPTA (Desvignes
et al.\ submitted), while following the observing restrictions
explained above. This results in some high-quality pulsars in crowded
areas of the sky being observed by less than 5 telescopes, or not
being included at all; this also means that some pulsars that are not
necessarily the best PTA sources are included in the list.

\section{Results}
\label{sec:results}

Processing of LEAP data is presently ongoing. During the second half
of 2014 the processing pipeline reached a level of maturity that
allowed us to transition to a scheme whereby the data of one epoch was
processed and analyzed before the data of the next epoch was
obtained. Here, we present results obtained from data from these
epochs, as well as data from a few specific epochs prior to the second
half of 2014, which have been processed during the development phase
of the pipeline.

\subsection{Coherence}
\begin{figure}
  \centering
  \includegraphics[angle=270,width=8cm]{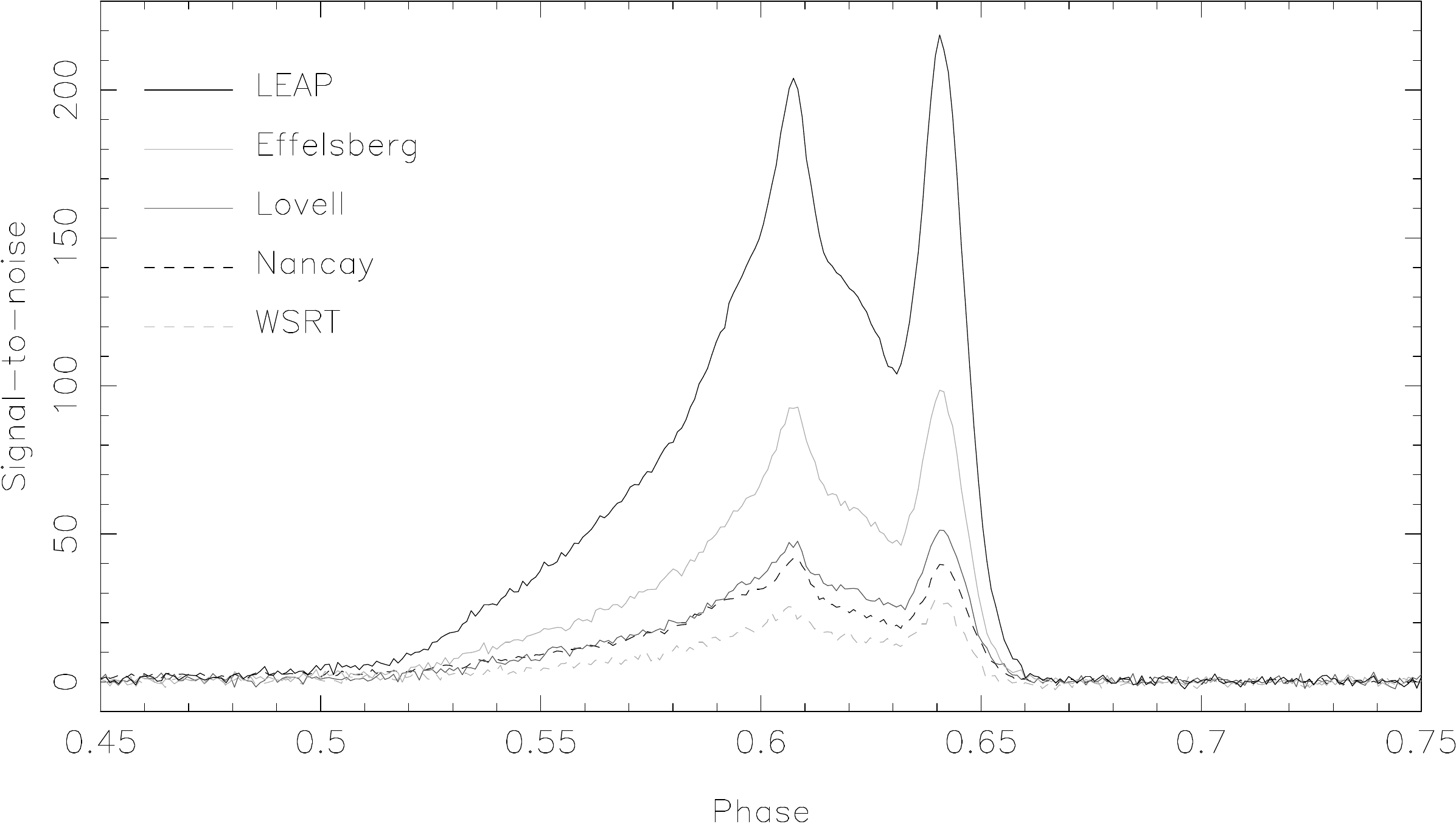}
  \caption{Pulsar profiles of PSR J1022+1001 from individual
    telescopes and their coherent addition, normalized based on their
    off-pulse rms. The raw data were obtained at MJD 56500, with an
    integration time of 30 minutes. The peak signal-to-noise ratios of
    Effelsberg, Jodrell Bank, Nan\c cay, WSRT and LEAP, are 97, 51,
    42, 30, 220, respectively, which corresponds to a near perfect
    coherency.}
  \label{fig:coherent}
\end{figure}

\begin{figure}
  \includegraphics[width=6cm,angle=-90]{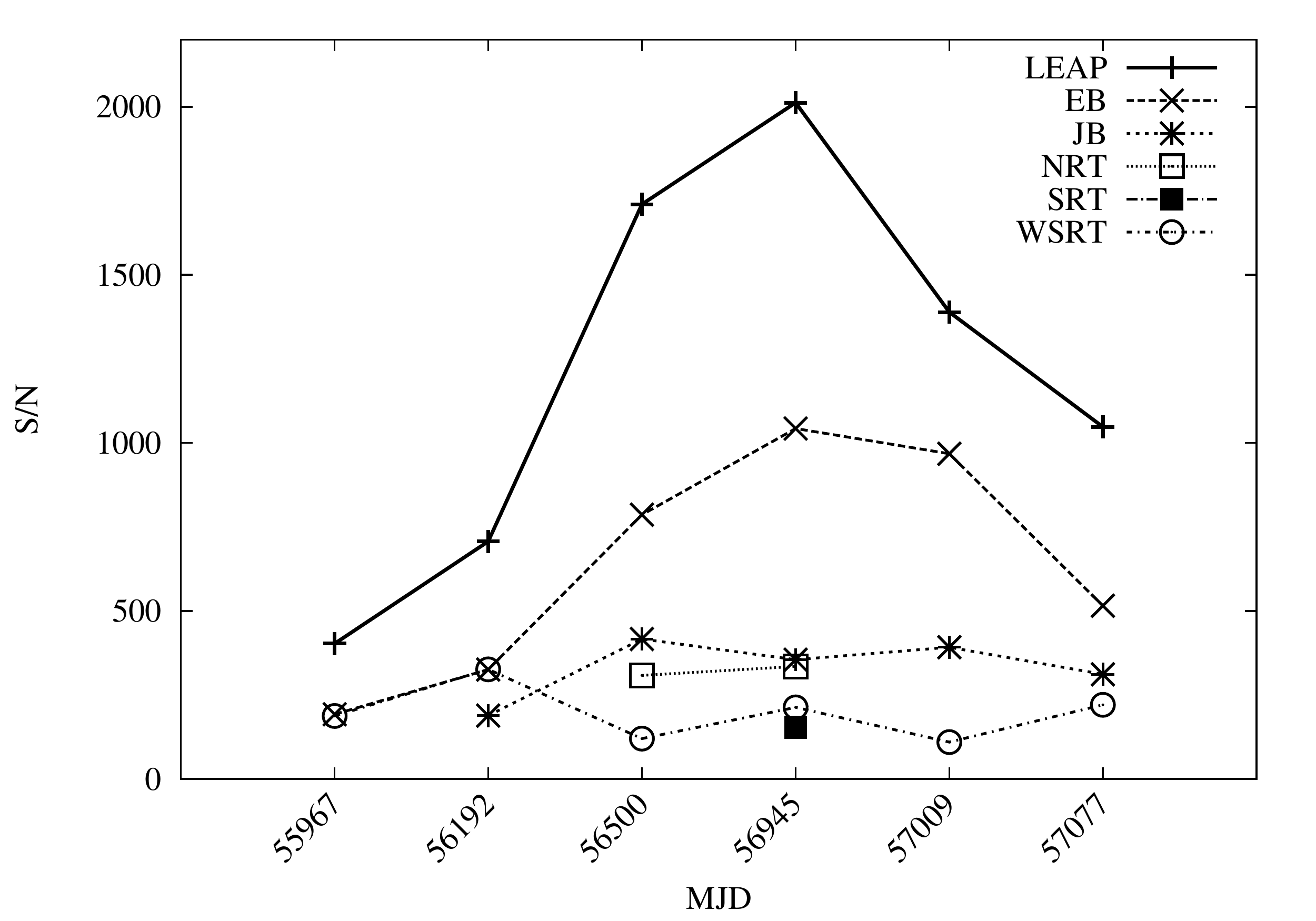}
  \caption{S/Ns from LEAP vs. S/Ns from the individual telescopes for
    PSR J1022+1001 for the observations where coherent addition could
    be performed. The earliest observation shown is from February
    2012, the last observation shown is from February 2015. The graph
    shows that the LEAP data provides the expected improvement in S/N,
    meaning that the sum of the S/Ns of the individual telescopes is
    roughly identical to the S/N of LEAP.}
  \label{fig:S/N}
\end{figure}

The correlation and addition of single-telescope baseband data using
the LEAP data reduction pipeline produces LEAP data with the expected
coherence. An example of such coherence is shown in
Fig.\,\ref{fig:coherent}, where we present the pulse profile of
PSR\,J1022+1001 from LEAP data compared to the profiles from
single-telescope data (all scaled to the off-pulse rms). In
Fig.\,\ref{fig:S/N} we present the S/Ns of all the LEAP profiles for
PSR J1022+1001, compared to those of the individual dishes, for the
months in which the pulsar signal was strong enough to perform
coherent addition. Full coherent addition is achieved when the
time series of the individual telescopes are perfectly in phase. The
LEAP S/N should then be similar to the sum of the S/Ns of the
individual telescopes. Fig.\,\ref{fig:coherent} and \ref{fig:S/N} show
that the S/N for LEAP is close to the sum of the S/Ns of the
individual telescopes, demonstrating that LEAP is achieving full
coherent addition when there is sufficient signal. Deviations from the
maximum S/N can be caused by an inaccurate fringe-solution (possibly
due to residual RFI or due to a non-linear phase-drift), or due to
improper polarization or amplitude calibration (see
Sect.\,\ref{sec:polcal} and \ref{sec:ampcal}).

With LEAP observing there are three possible data combinations. The
most sensitive of these is clearly when we combine all the dishes
involved coherently over the full LEAP bandwidth. In the few cases
where coherent addition is not possible the incoherent sum of the
available dishes, over the LEAP bandwidth, gives us the best
sensitivity. This assumes that a sufficient number of dishes
(i.e.\ more than 2) is involved in the sum. Otherwise, the incoherent
combination of the TOAs, as opposed to the raw data, from the wide
bandwidth observations from the individual telescopes is used. This is
because the sensitivity of the incoherent sum scales as the
square-root of the number of dishes while the sensitivity of the
combination of the TOAs determined from the wide-band data scales as
the square-root of the ratio of the bandwidth available to the dishes
over that available to LEAP. In all cases we end up with a better
result for the overall sensitivity compared to what would be possible
with a single telescope observation from one of the LEAP dishes.

Based on the LEAP observations that have been fully processed at the
time of submission of this paper, 51\% of the sources were processed
coherently with more than 80\% coherency, 8\% were processed
coherently with 60 to 80\% coherency, and the remaining 41\% were
processed incoherently. The reasons for the poor coherency achieved
for some of the pulsars are a combination of poor S/N due to
scintillation, imperfect polarization calibration, large or non-linear
fringe drifts due to ionospheric conditions or the Nan\c cay clock,
and RFI across the LEAP band.

While these coherence numbers are lower than hoped, we have already
improved our polarization calibration routines and our RFI mitigation
procedures as described elsewhere in the paper, therefore these
statistics are already improving\footnote{The large data sizes
  involved here meant that previously combined data could not be
  reprocessed with these improvements as the LEAP combination was
  already done and the individual telescope data deleted.}. As
discussed above, even if full coherence is not achieved, the various
forms of incoherent combination already result in significant improved
TOA precision compared to an observation with a single EPTA
telescope. However we also do see ways to improve our ability to
achieve coherence more often and discuss some of them here.  When
using the pulsar for fringe-finding we use pulsar gating, that is we
use only the on-pulse region to improve the S/N, to further improve
this we will subtract the off-pulse region which can improve
sensitivity in regions of the sky where there might be bright sources
in the field-of-view of one or more of the telescopes. We will also
implement a new algorithm for identifying the on-pulse region when the
pulsar has low S/N which will use the predicted phase of the pulse and
a template profile. The long baselines mean that ionospheric
conditions can lead to significant phase drift as a function of
frequency, as can the less stable clock at Nan\c cay. One way to
overcome this is to implement a more sophisticated fringe-fitting
routine which searches over a range of fringe-drift rates to look for
the best drift rate to maximise the S/N of the fringe detection
without having to go to too short integration times. Another option we
are investigating for the near future is to increase the bandwidth
used for LEAP. Not only does this lead to a higher S/N through the
increased bandwidth, it also increases the chance of detecting the
pulsar when it scintillates over a bandwidth smaller than the observed
bandwidth. This improves our chances of getting coherent solution in
that part of the band, but the delays can also be used to search for
fringes where the signal is weaker. So overall the prospects are good
for significantly improving the coherence that can be achieved for
LEAP.

\subsection{Improvement in timing accuracy}
\begin{figure}
\centering
  \includegraphics[angle=270,width=8cm]{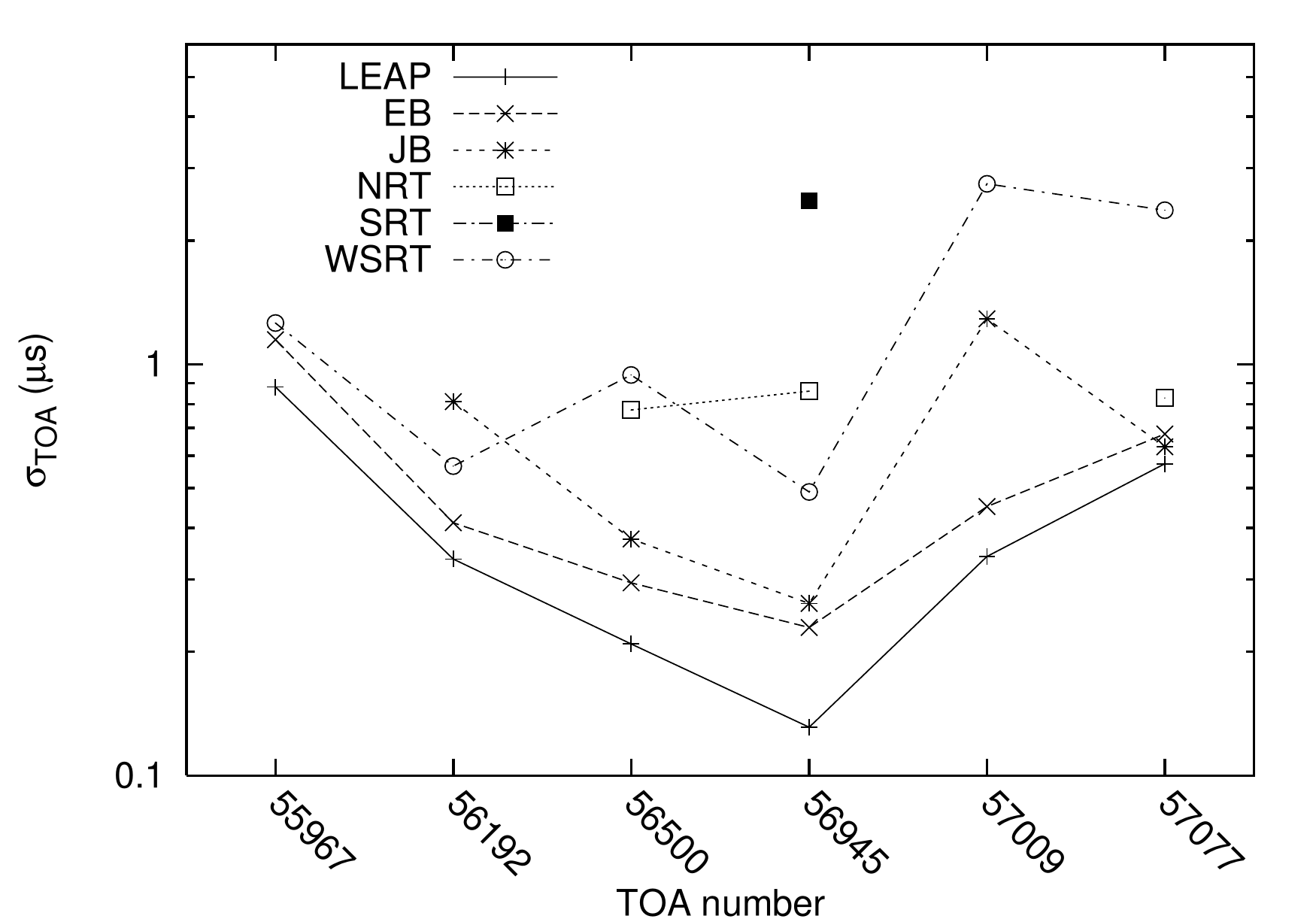}
  \caption{TOA uncertainties from LEAP for PSR J1022+1001 compared
    with those obtained from single-telescope data, which were
    acquired simultaneously but with broader bandwidth.  The full
    ordinary bandwidths of Jodrell Bank and Nan\c cay are 400 and 512\,MHz,
    respectively.  The TOA uncertainties from Effelsberg and
    WSRT were extrapolated from 128\,MHz to 200 and 160\,MHz,
    respectively (these are the bandwidths used in the ordinary
    on-site EPTA timing campaigns).  The available EPTA timing
    bandwidth at SRT is currently the same as LEAP.}
  \label{fig:sigma_comp}
\end{figure}

The LEAP coherent addition makes optimal use of the acquired radio
signals from each individual telescope. At present it uses a smaller
bandwidth than in ordinary EPTA timing observations at most
telescopes. This is in part due to the limited bandwidth available
with PuMa\,II at the WSRT, but also due to current limitations on data
rates and data storage. In the future we plan to expand the bandwidth
observed with LEAP. To demonstrate that LEAP can improve the data
quality, as compared to the individual telescope observations with
wider bandwidth, we compare the LEAP TOAs of PSR~J1022+1001 with those
from single telescopes (see Fig.~\ref{fig:sigma_comp}). The TOAs from
Jodrell Bank and Nan\c cay were derived directly from the simultaneous
observations in ordinary timing mode, with bandwidths of 400 and
512\,MHz, respectively, while SRT TOAs are limited to the LEAP
bandwidth (128 MHz). The TOA uncertainties from Effelsberg and WSRT
were extrapolated based on the LEAP bandwidth to 200\,MHz and
160\,MHz, respectively, since data acquisition with a wider bandwidth
is not feasible at these telescopes during LEAP observations. It can
be seen that compared with regular timing observations at the
individual telescopes, the TOAs obtained from coherently-added LEAP
data have smaller uncertainties. This is even more striking when one
considers that the observations at Jodrell Bank and Nan\c cay observe
over the same full 400 and 512\,MHz bandwidth as regular timing
observations. The bands that are not used for coherent addition are
dedispersed and folded as if they were regular timing observations,
hence contributing to the timing dataset of those particular
telescopes. Therefore, observations in LEAP mode clearly improve the
sensitivity compared to the individual telescopes, as expected.

\begin{figure}
\centering
  \includegraphics[angle=270,width=8cm]{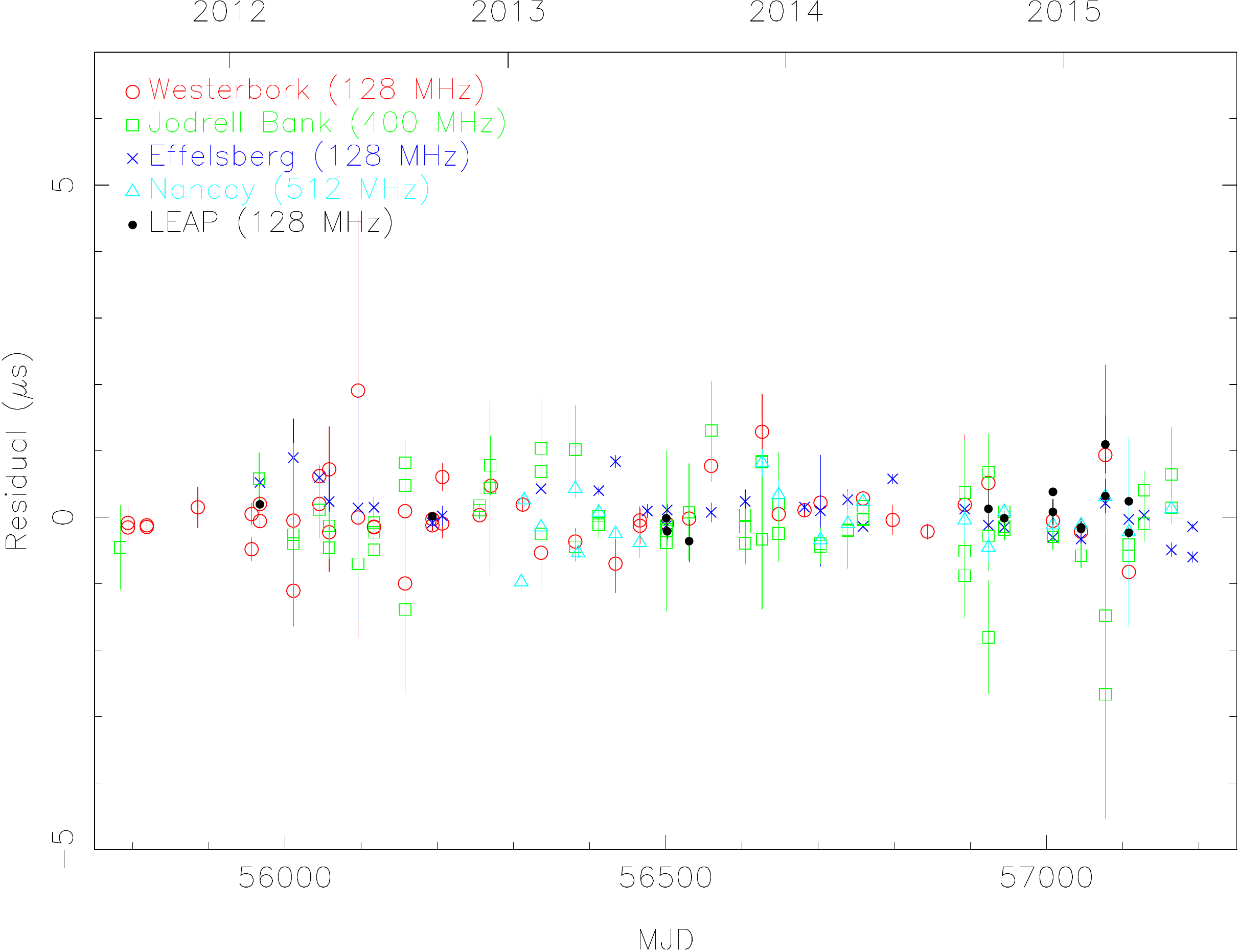}
  \caption{Timing residuals of PSR~J1713+0747 obtained from single
    telescope data (colored points), as well as the coherently added
    LEAP data (black points). These residuals are computed by
    comparing the TOAs against the long-term EPTA timing solution of
    PSR~J1713+0747 (Desvignes et al.\,submitted). No parameters, other
    than timing offsets between the telescopes, were fitted for. Over
    this five year timespan the data from the individual telescopes
    participating in LEAP, as well as the coherently added LEAP data
    presently available, allow the timing solution to be constrained
    to an rms of 0.25\,$\mu$s. The solution using only TOAs determined
    from the coherently added LEAP data has an rms of
    0.18\,$\mu$s. For the Jodrell Bank and Nan\c cay telescopes the
    TOAs from the data obtained over the full instrument bandwidth are
    shown.  }
  \label{fig:j1713_residuals}
\end{figure}

Furthermore, in Fig.\,\ref{fig:j1713_residuals} we compare the TOAs of
PSR~J1713+0747 determined from both the individual telescope data as
described above, as well as the LEAP coherent sum, with the long-term
EPTA timing solution (Desvignes et al.\,submitted). This timing
solution is based on data from the individual telescope participating
in the LEAP project (Effelsberg, Jodrell Bank, Nan\c cay and WSRT) and
obtained over a 17.7\,year long timespan between October\,1996 and
June\,2014. The data for the long-term EPTA timing solution were
obtained with older generation instruments. No parameters in the
timing solution were fitted for except for timing offsets between the
individual telescopes. Fitting only for these timing offsets yields a
solution with an rms of 0.25\,$\mu$s when using TOAs from both the
individual and coherently added LEAP data spanning nearly
4\,years. Using only TOAs from the coherently added LEAP data improves
the rms to 0.18\,$\mu$s. For comparison, the long-term EPTA timing
solution has an rms residual of 0.68\,$\mu$s over the 17.7\,year
observing span (Desvignes et al.\,submitted). The TOAs determined from
individual telescope data significantly improve the timing precision,
primarily due to the use of a new generation of instruments, capable
of coherent dedispersion over larger bandwidths. The TOAs determined
from coherently combined LEAP data provide a further improvement on
top of that.

\subsection{Phase jitter and single pulse studies}
\begin{figure}
  \includegraphics[angle=270,width=8cm]{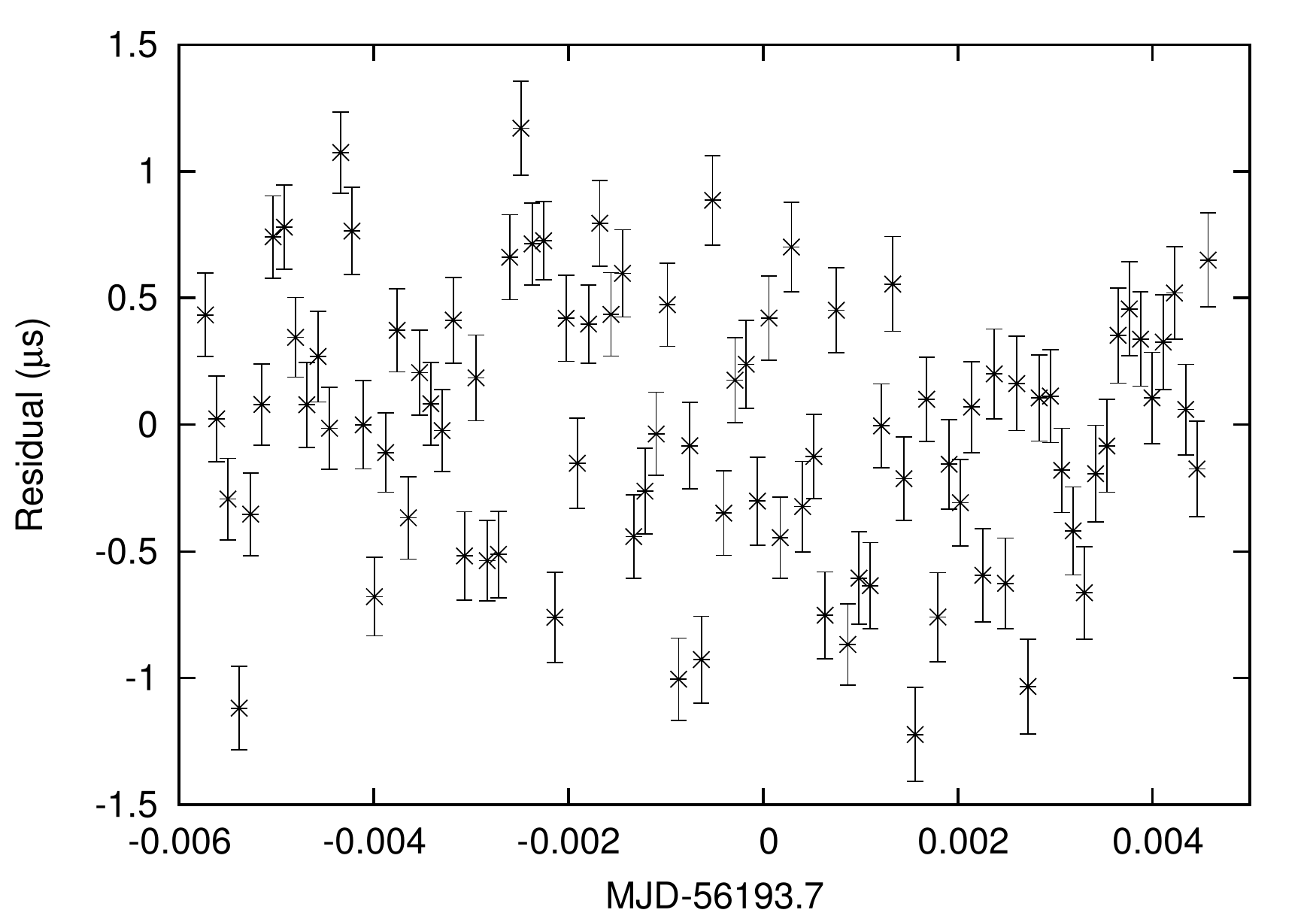}
  \caption{Timing residuals of PSR~J1713+0747 over a period of 15 min,
    for each 10-s integration. The observations were performed on MJD
    56193, included Effelsberg, Nan\c cay, and WSRT, and the
    generation of the LEAP data achieved a coherency of 95\%.  The rms
    residual is 522\,ns and the corresponding reduced $\chi^2$ is
    9.47.}\label{fig:jitter}
\end{figure}

\begin{figure}
  \includegraphics[angle=270,width=8cm]{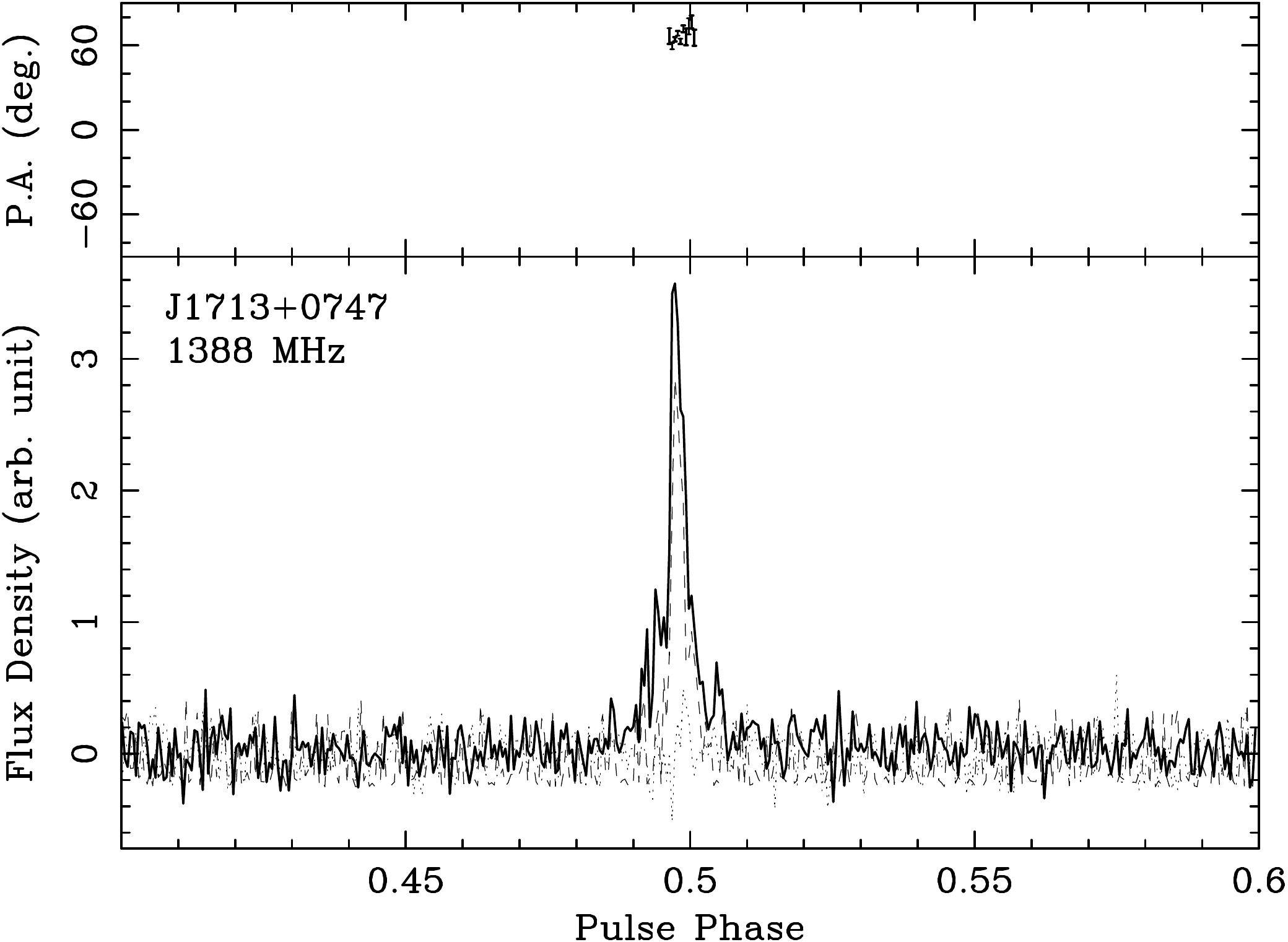}
  \caption{Polarization profile of a single pulse from PSR~J1713+0747,
    obtained from the observation used in
    Fig.~\ref{fig:jitter}.}\label{fig:sglsamp}
\end{figure}

LEAP delivers a sensitivity that is rivaled only by Arecibo, the
largest single-dish radio telescope on Earth. The data are therefore
ideal for studies of the phase jitter of integrated profiles and
single pulses of MSPs, which are not often feasible with
single-telescope data due to low S/N.  Fig.~\ref{fig:jitter} shows an
example of such an analysis for PSR~J1713+0747.  The observations were
carried out with Effelsberg, Nan\c cay, and WSRT on MJD~56193.  The
plot shows timing residuals for 10-s integrations for a 15-min
observing time. The TOA errors corresponding to measurement
uncertainties due to radiometer noise were estimated by the classic
template matching method \citep{tay92}. To calculate the residuals, we
used the ephemeris from the EPTA timing release (Desvignes et
al.\,submitted) without fitting for any parameters.  We see that the
error bars clearly underestimated the scatter of the residuals, which
is an indicator of phase jitter \citep[e.g.][]{lvk+11}. The rms
residual is 522\,ns with a reduced $\chi^2$ of 9.47. Following the
method in \cite{lkl+11}, this leads to an estimated jitter noise of
494\,ns for a 10-s integration time.  This is consistent with
previously published results \citep{sc12,dlc+14}. From the
coherently-added LEAP data, we also managed to obtain single pulses of
the pulsar with fully-calibrated polarization at a time resolution of
2.2\,$\mu$s, an example of which can be found in
Fig.~\ref{fig:sglsamp}. The single pulses have sharp features and
significant linear polarizations. Further investigation of the single
pulses from PSR J1713+0747 will be presented in a separate paper.

\subsection{Pulsar searching}
The increased sensitivity of the LEAP tied array allows searches for
weak pulsars with known positions. Though the LEAP tied array beam of
the full LEAP array is small, beamforming can be used to tile out the
incoherent beam.

As a proof of concept, we have performed a blind search on
5\,min of coherently-added LEAP data of the double neutron star
PSR\,J1518+4904, with the aim of detecting pulsations from the second
neutron star. The baseband data were acquired at MJD 56193 with
Effelsberg and the WSRT, and were later combined with nearly full
coherency. The resulting Nyquist-sampled timeseries of each 16-MHz
subband were then used to form a filterbank file with 1-MHz
channels. Next we combined the filterbank files from each individual
subband to yield the full observing bandwidth and used the
\textit{PRESTO} software package to search for pulsations.

In total, 33 candidates were detected with the same DM as
PSR\,J1518+4904, all of which were harmonics of the pulsar or
attributed to RFI. No pulsations with a non-harmonic period were found
from an initial investigation down to a flux limit of 0.31\,mJy. As
PSR\,J1518+4904 is part of the monthly LEAP observing sessions, we
will be able to use all coherently-combined data on this system for
the most sensitive search to date for radio emission from its neutron
star companion.

\section{Conclusions and prospects}
\label{sec:conclusions} In this paper we present an overview of
the LEAP project, which coherently combines data from up to five 100-m
class radio telescopes in Europe, forming a tied-array telescope. We
observe a subset of the EPTA MSPs with a sensitivity that cannot be
achieved by the individual participating telescopes. The LEAP project
emerges as a natural result of the many years of collaboration between
the EPTA groups.  Instead of merely sharing their TOAs for GW
detection purposes, the EPTA telescopes in the LEAP project are
combined using VLBI techniques to form a fully-steerable 195-m
equivalent dish, forming one of the most sensitive pulsar observation
instruments to date.

We describe the LEAP setup and operation, starting from the data
acquisition setup at the participating telescopes, the transfer of
data to the centralized LEAP computing infrastructure at Jodrell Bank,
to the final processing of the monthly LEAP observing runs. We have
also presented the main characteristics of the pipeline that was
developed for the processing the data. We describe the challenges of
achieving high timing precision, in great part due to the many
differences in the telescopes and their pulsar observing
systems. These differences were managed either fully in software
(incorporated into the LEAP pipeline), or with hardware upgrades when
these were inevitable. The development of our own end-to-end pipeline
(individual telescope data, polarization calibration, RFI mitigation,
correlator and tied-array adder) not only provided us with the
flexibility to overcome all of these obstacles, but also allowed us to
take the most out of each telescope. The efforts placed into making
LEAP a reality have however been rewarded by the quality of the
results. As we have shown, the coherency of the added individual
telescope data can reach 100\%. In addition, the TOA uncertainty of
the LEAP data is less than that of the individual telescopes, even
though the LEAP bandwidth is a few times smaller.

Although the main aim of LEAP is to provide high precision pulsar
timing data towards a direct detection of GWs, its high sensitivity
and flexibility as an observing system enable it to go beyond this
scope and pursue broader pulsar-related science. Pulse phase jitter
and single pulse studies, which are demanding in terms of sensitivity,
are ideal for LEAP. This was best demonstrated with the single pulse
detections of PSR\,J1713+0747 during one of the standard LEAP
observations. We have also demonstrated that LEAP is capable of
performing targeted pulsar searches in a case study using
PSR\,J1518+4904. Even though its current operation mode does not allow
it to be used as a generic pulsar searching instrument, its high
sensitivity makes it a perfect tool for investigating known binaries
and looking for pulsations from pulsar companions in order to identify
double-pulsar systems.  Moreover, LEAP has recently been used to
observe the Galactic center magnetar PSR\,J1745$-$2900 at frequencies
higher than used in the typical LEAP runs, in order to determine the
scattering properties of the ISM towards the Galactic centre. This
study used VLBI imaging techniques and helped define the best search
strategies for pulsars close to Sgr\,A$^*$ \citep{wuc15}.

The addition of LEAP data to the current PTA data sets will
significantly improve PTA data quality. We are currently finalizing
the LEAP timing data set of the data obtained to date, and will use
these data to perform a search for GWs and place upper limits on the
GW amplitude. We can already extrapolate the results of our currently
processed data to the full time span of 3.2 years, by counting the
number of telescopes that joined each observing session. Assuming 90\%
coherency and using the red noise parameters of each pulsar measured
from the much longer EPTA data set, we can calculate the statistics of
the expected timing noise and measurement accuracy, then derive upper
limits on the amplitude of the GW background using a Cramer-Rao
bound. For a spectral index of $-2/3$ (i.e. a stochastic GW background
dominated by supermassive binary black holes), the LEAP upper limit on
the dimensionless strain amplitude $A_\mathrm{c}$ is $A_{\rm c} ({\rm
  1 yr^{-1}})\le 1.2\times 10^{-14}$, using extrapolated data of four
LEAP pulsars, PSR\,J0613$-$0200, J1022$+$1001, J1600$-$3053, and
J1713$+$0747.  With only 3.2 years of data, such an upper limit is a
factor 2 to 5 higher compared to the published results that used
10-year long data sets and more pulsars
\citep{hlj+11,dfg+13,src+13,ltm+15}.

Dedicated funding for the LEAP project officially ended in September
2014. However, the unique character and the success of LEAP have
justified its continuation at all participating telescopes, which have
provided the necessary monthly observation time. While this paper
provides an overview of the LEAP project, several papers are presently
in preparation that provide details of the instrumentation, pipeline
and the calibration, as well as present results from the LEAP project.

\section*{Acknowledgements}
The European Pulsar Timing Array (EPTA) is a collaboration of European
institutes to work towards the direct detection of low-frequency
gravitational waves and to implement the Large European Array for
Pulsars (LEAP). The authors acknowledge the support of the colleagues
in the EPTA. In particular, we would like to thank the following
colleagues for their great help with our instrumentation, observations
and administrative issues (in alphabetical order): I.~Cognard,
G.~Desvignes, B.~Dickson, A.~Holloway, C.~Jordan, P.~Lespagnol,
A.~G.~Lyne, P.~Stanway, G.~Theureau, and WeiWei Zhu. In Sardinia we
would like to thank the SRT Astronomical Validation Team, and in
particular: M. Burgay, S. Casu, R. Concu, A. Corongiu, E. Egron,
N. Iacolina, A. Melis, A. Pellizzoni and A. Trois. We are grateful to
A.~Deller, M.~Kettenis and Z.~Paragi for sharing their knowledge of
VLBI. The work reported in this paper has been funded by the ERC
Advanced Grant ``LEAP'', Grant Agreement Number 227947 (PI
M.~Kramer). KJL acknowledge support from National Basic Research
Program of China, 973 Program, 2015CB857101 and NSFC 11373011.  The
100-m Effelsberg Radio Telescope is operated by the
Max-Planck-Institut f\"{u}r Radioastronomie at Effelsberg. The
Westerbork Synthesis Radio Telescope is operated by the Netherlands
Institute for Radio Astronomy (ASTRON) with support from The
Netherlands Foundation for Scientific Research (NWO). The Nan\c{c}ay
Radio Observatory is operated by the Paris Observatory, associated
with the French Centre National de la Recherche Scientifique. Pulsar
research at the Jodrell Bank Centre for Astrophysics and the
observations using the Lovell Telescope are supported by a
consolidated grant from the STFC in the UK. The Sardinia Radio
Telescope is operated by the Istituto Nazionale di Astrofisica (INAF)
and is currently undergoing its astronomical validation phase.

\bibliographystyle{mnras}

\bsp

\label{lastpage}

\end{document}